\documentclass[acmsmall]{acmart}

\usepackage[utf8]{inputenc} %
\usepackage[T1]{fontenc}

\usepackage{makecell}
\usepackage{multirow}
\usepackage{booktabs}
\usepackage{tabularx}
\usepackage{arydshln}
\usepackage{todonotes}
\usepackage[group-separator={,}]{siunitx}
\usepackage[labelformat=simple]{subcaption}

\usepackage{xcolor}
\usetikzlibrary{patterns,decorations.pathreplacing,arrows.meta,shapes.symbols,shapes.geometric}
\usepackage{rotating}

\usepackage{wrapfig}

\usepackage[absolute,showboxes]{textpos}

\setcopyright{cc}
\setcctype{by}
\acmJournal{PACMNET}
\acmYear{2025} \acmVolume{3} \acmNumber{CoNEXT3} \acmArticle{15}
\acmMonth{9} \acmPrice{}\acmDOI{10.1145/3749215}



\received{December 2024}
\received[revised]{June 2025}
\received[accepted]{June 2025}

\newcommand{\result}[1]{}

\definecolor{myred}{cmyk}{0, 0.7808, 0.4429, 0.1412}

\newcommand{\done}[1]{}

\definecolor{positive}{HTML}{3C8031}
\definecolor{negative}{HTML}{AF3235} %
\definecolor{neutral}{HTML}{E1A205} %

\usepackage{pifont}
\newcommand{\cmark}{\ding{51}}%
\newcommand{\xmark}{\ding{56}}%

\usepackage{xspace}
\newcommand{\etal}{\textit{et al.}~}
\newcommand{\eg}{\textit{e.g.,}~}
\newcommand{\ie}{\textit{i.e.,}~}

\newcommand{\cf}{\textit{cf.,}~}
\newcommand{\one}{({\em i})\xspace}
\newcommand{\two}{({\em ii})\xspace}
\newcommand{\three}{({\em iii})\xspace}
\newcommand{\four}{({\em iv})\xspace}
\newcommand{\five}{({\em v})\xspace}

\makeatletter
\renewcommand{\paragraph}[1]{\vspace*{0.03in}\noindent{\bf #1.}\hspace{0.25ex \@plus1ex \@minus.2ex}}
\newcommand{\paragraphNoDot}[1]{\vspace*{0.03in}\noindent{\bf #1}\hspace{0.25ex \@plus1ex \@minus.2ex}}
\makeatother

\begin{document}

\title{Analysis and Characterization of Scanner Activities in the IPv6}
\title[A Detailed Measurement View on IPv6 Scanners and Their Adaption to BGP Signals]{A Detailed Measurement View on IPv6 Scanners and\\ Their Adaption to BGP Signals}

\author{Isabell Egloff}
\orcid{0009-0003-5041-050X}
\affiliation{%
  \institution{HAW Hamburg}
  \city{Hamburg}
  \country{Germany}
}
\email{isabell.egloff@haw-hamburg.de}

\author{Raphael Hiesgen}
\orcid{0000-0002-1676-8108}
\affiliation{%
  \institution{HAW Hamburg}
  \city{Hamburg}
  \country{Germany}}
\email{raphael.hiesgen@haw-hamburg.de}

\author{Maynard Koch}
\orcid{0009-0009-3698-1342}
\affiliation{%
  \institution{TU Dresden}
  \city{Dresden}
  \country{Germany}
}
\email{maynard.koch@tu-dresden.de}

\author{Thomas C. Schmidt}
\orcid{0000-0002-0956-7885}
\affiliation{%
  \institution{HAW Hamburg}
  \city{Hamburg}
  \country{Germany}}
\email{t.schmidt@haw-hamburg.de}

\author{Matthias W\"ahlisch}
\orcid{0000-0002-3825-2807}
\affiliation{%
  \institution{TU Dresden}
  \city{Dresden}
  \country{Germany}
}
\email{m.waehlisch@tu-dresden.de}

\definecolor{boxgray}{rgb}{0.93,0.93,0.93}
 \textblockcolor{boxgray}
 \setlength{\TPboxrulesize}{0.7pt}
 \setlength{\TPHorizModule}{\paperwidth}
 \setlength{\TPVertModule}{\paperheight}
 \TPMargin{5pt}
 \begin{textblock}{0.8}(0.1,0.04)
   \noindent
   \footnotesize
   If you refer to this paper, please cite the peer-reviewed publication:  
   Isabell Egloff, Raphael Hiesgen, Maynard Koch, Thomas C. Schmidt, and Matthias Wählisch. 2025.
   A Detailed Measurement View on IPv6 Scanners and Their Adaption to BGP Signals.
   \emph{Proceedings of the ACM on Networking (PACMNET) 3, CoNEXT3 (September 2025), 15:1–15:23}. https://doi.org/10.1145/3749215
\end{textblock}

\begin{abstract}
Scanners are daily visitors of public IPv4 hosts. Scanning IPv6 nodes successfully is still a challenge, which an increasing crowd of actors tries to master. In this paper, we analyze IPv6 scanning under various network conditions to disclose the impact on scanning. We deploy four IPv6~network telescopes, including a reactive /48~telescope and a proactive /32 telescope that is periodically reconfigured by changing BGP announcements.
We provide a longitudinal study of eleven months and classify the observed scanners w.r.t. their temporal behavior, their target and network  selection strategies, as well as their  individual tools, fingerprints, and correlations across categories. We find that silent subnets of larger covering prefixes remain mainly invisible, whereas BGP prefix announcements quickly attract attention by scanners. Based on our findings, we derive operational guidance on how to deploy network telescopes to increase visibility for IPv6 scanners and understand corresponding biases.
\end{abstract}

\maketitle

\section{Introduction}
\label{sec:intro}

\begin{wrapfigure}{r}{0.4\textwidth}
	\centering
	\vspace{-15pt}
	\resizebox{0.4\textwidth}{!}{\input{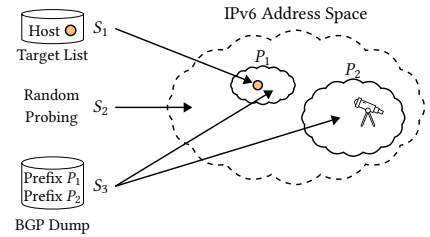}}
	\caption{IPv6 address probing methods.}
	\label{fig:intro:setup2}
\end{wrapfigure}

Scanning is prevalent on the Internet~\cite{apt-bhs-07,dbh-ivis-14}. Researchers, commercial services, as well as malicious actors explore the IPv4 address spaces regularly and with high intensity. Many of these build their tooling on stateless scanning~\cite{dwh-zfiws-13,hnkds-sunws-22}, which allows for traversing all IPv4~addresses within less than an hour. The huge IPv6 address space renders this impossible. It is thus vital for research, operations, and security to develop methods for observing and understanding the emerging ecosystem of IPv6 scanners. 

Scanning IPv6 nodes means mastering the challenge of finding them. Different strategies have been developed (see \autoref{fig:intro:setup2}), more are expected to emerge. Simple random probing, such as $S_2$ in  \autoref{fig:intro:setup2}, does not scale for an Interface Identifier (IID) space of $2^{64}$ addresses per subnet. Nevertheless, random probing can be successful if its distribution accounts for IPv6-inherent address structures. Scanners can also execute along a target list ($S_1$), which may be taken from some static hitlist or generated algorithmically on the fly. Scanners can also react to external events ($S_3$), such as the publication of DNS records or---more network-centric---the announcement of new prefixes via BGP.

Measuring  the IPv6 scanning landscape and obtaining a meaningful picture is about as challenging as the scanning itself.  An operator of a public IPv6 network  without any further attractors may experience close to no scan visits. This challenges the operation of network telescopes (and honeypots), which are the  traditional observatories of scanners.
 Deploying specific attractors to influence scanning will increase packet reception, but also introduces biases that misguide conclusions.  

Internet scanning regularly precedes malicious attacks. Measuring malicious scanners may therefore serve as a quantitative assessment of the current threat landscape~\cite{hnbkh-adeci-24,kocei-lelwd-24,hnsw-licms-24}. Accurate measurements and quantitative assessments, however, are intricate. Distributed honeypots have shown to span a very selective view on the Internet~\cite{nkkhs-advmd-23}. Operating an IPv4 telescope and interpreting its data correctly likewise exhibits high complexity~\cite{mmcgm-lloln-25}. Soundly assessing the limitations and biases of our measurement instruments is essential but difficult. In IPv6, these difficulties exponentiate, since any scope of perception remains tiny compared to the huge address space. In view of this, the importance of understanding the interplay between network characteristics and scanner reactions becomes even more evident.

In this paper, we explore the reactivity of scanners to network embeddings of telescopes by performing a large measurement study, in which we deploy four  telescopes with different network attachments for eleven months. With this setup, we want to study the effective adaptation of scan traffic to these telescopes, in particular to changes in BGP.
After summarizing the state of the art on IPv6 address characteristics and IPv6 scan studies (\S2), we present our main contributions:

\begin{enumerate}
	\item In controlled experiments, we observe IPv6 scanning during 11 months from four network telescopes with contrasting network properties.
	This includes a proactive telescope that varies visibility in BGP with up to 17 IPv6~prefixes (\S\ref{sec:method}).
	We capture and analyze 51M packets ~(\S\ref{sec:traffic}).

  \item We derive a taxonomy and classify scanners following a thorough analysis of their temporal behavior, of their target selection strategies, and their origins (\S\ref{sec:taxonomy}).
  
  \item We analyze and compare all four telescopes during a 12 week observation period, revealing unique traffic patterns in terms of packet intensity, sources, and scanning strategies (\S\ref{sec:behavior}).

  \item We analyze the reactions of scanners to our 8 months of changing BGP signals and find a strong correlation with packet arrival (+286\%). 70\% of all scanners were observed only once, 
9\%  target uniformly all of the announced prefixes and account for 63\% of all packets~(\S\ref{sec:splitting}).

  \item We derive operational guidance on the visibility and expected biases of IPv6 telescopes (\S\ref{sec:discussion}). 
\end{enumerate}

\section{Background and Related Work}
\label{sec:background}

The huge IPv6 address space challenges Internet measurements.
Discovering active endpoints is in focus of of active research, industry projects, and malicious actors.
Notably, TU Munich maintains a comprehensive hitlist of responsive IPv6 hosts~\cite{gsgc-siitc-16,zssgc-rcdir-22}, which estimates a lower bound  of all active IPv6 addresses.
While scanners actively explore the address space, incoming scan activities can be observed through telescopes to analyze scanners.  \autoref{tbl:rw:overview} summarizes related scientific work.

\paragraph{IPv6 address characteristics}
Manual configuration often leads to four easy-to-predict and memorable patterns~\cite{RFC-7707}.
\one~Low-byte addresses use Interface Identifier (IID) bytes of zero except for the least significant byte, \eg \texttt{2001:db8::1},
\two service port addresses embed the port of a running service in the IID, \eg \texttt{2001:db8::443} for HTTPS,
\three IPv4-based addresses embed the IPv4 address of the network interface in the IID, \eg \texttt{2001:db8::192.0.0.1}, and
\four wordy addresses include semantic hints such as \texttt{2001:db8::cafe}.

Stateless address autoconfiguration~\cite{RFC-4862} maps MAC addresses and thus also follows predictable patterns in its assignment strategy if not used with a privacy extension~\cite{RFC-4941}.
One strategy is to scan these address ranges first, as the density of active hosts is expected to be higher. %
When observing scanners in this work, we analyze target address patterns to infer their scanning~strategies.

\paragraph{IPv6 scanning strategies}
Early drafts of RFC~7707~\cite{RFC-7707} dating back to 2012 already described active IPv6 address discovery strategies.
Ullrich \etal~\cite{ukkw-ripsa-15} were the first to publish on pattern-based scanning techniques, followed by Foremski \etal~\cite{fpb-eiusi-16} who probabilistically modeled IPv6 addresses. %
Starting in 2019, multiple studies appeared on target generation algorithms (TGA)~\cite{zf-6mcbh-24,cslhl-6aiad-23,ljhhc-6rlta-23,ll-6tiag-23,hcwyz-6hdii-23,yhcwz-6gaap-22,ychz-6elat-22,hcwsx-6rlat-21,cgxlf-6imtg-21,cxgsx-6lmvs-20,cgx-6gcva-20,sywhl-deepi-22,lxlxz-6edda-19}, which can be roughly divided into \one static and \two dynamic TGAs~\cite{skzcg-taetg-23}.
While static algorithms only generate potential candidates for scanning based on a fixed training set, dynamic TGAs adjust their training set by evaluating the activity of generated addresses immediately through active scanning.
In our scanning analysis, we use the order of selected target addresses to infer scanning~strategies.

\paragraph{Passive network telescopes}
Ford \etal~\cite{fsr-irid-06} were the first to analyze IPv6 scanners by examining the background radiation of a /48 prefix in 2005 and 2006.
They only received 12~packets.
Seventeen years later, Ronan \etal~\cite{rm-rrint-23} re-examined the same /48 prefix.
They collected over 5k packets in 6~months. %
Most sources sent ICMPv6, about a fifth TCP, and a few UDP.
Scanners generally iterated the network prefix systematically.
In 2021, Liu \etal~\cite{lhlbl-intnt-21} investigated the background radiation of a previously unused /20 IPv6 prefix and collected more than 2.9M packets during roughly 6~months.
The authors observed similar shares of transport protocols.
95\% of the packets originating from only 10 sources.
Our measurements also use passive telescopes but of controlled, variable BGP~exposure.

\setlength{\tabcolsep}{2.75pt}
\begin{table*}%
  \scriptsize
  \centering
\caption{Overview of related work that observes IPv6 scanners. In contrast to prior work, we perform an active and controlled experiment on the control plane and deploy several prefixes of different characteristics for comparison. \emph{Passive} telescopes originate no packets. \emph{Traceable} telescopes originate or receive traffic controlled by the authors. \emph{Active} telescopes react to connection attempts. Duration is some weeks (w) or months (m).}
  \label{tbl:rw:overview}
  \begin{tabular}{lrrrcccrrr}
    \toprule
				       & \multicolumn{2}{c}{Time} & \multicolumn{4}{c}{Telescope} & \multirow{2.7}{*}{\makecell[c]{Announced \\ Prefix Sizes}} & \multirow{2.7}{*}{\makecell[c]{Application\\Attractors}}  & \multirow{2.7}{*}{\shortstack{Packets\\Received}}                                                                                                                                          \\
    \cmidrule(lr){2-3}
    \cmidrule(lr){4-7}
	  Publication                         & Year    & Duration   & Size & Passive & Traceable & Active &              &  & \\
    \midrule
	  Ford \etal~\cite{fsr-irid-06}       & '04-'06 & 16m        & /48  & \cmark  & \xmark   & \xmark & /48 &  & 12   \\
	  Czyz \etal~\cite{clmbk-uiibr-13}    & '12-'13     & 3m         & 5$\times$/12 & \cmark  & \xmark & \xmark & 5$\times$ /12 &     & 209M \\
	  Fukuda \etal~\cite{fh-wkidd-18}     & '17-'18     & 9m         & /37  & \cmark  & \xmark   & \xmark & /37 &            & 15k \\
	  Strowes \etal~\cite{swosf-dawvn-20} & '20     & 1w         & /12 & \xmark  & \cmark & \xmark & /12, 4$\times$/32, 4$\times$/48 &  & 6.5M \\
	  Liu \etal~\cite{lhlbl-intnt-21}     & '19-'21     & 4m, 3w, 3w & /20  & \cmark  & \xmark   & \xmark & /20 &            & 2.9M \\
	  Tanveer \etal~\cite{tspn-gduia-23}  & '21     & 48w        & /56, 24$\times$/64 & \xmark  & \cmark & \xmark & (unclear)   & Web, DNS, NTP, TOR & 14.6M \\
	  Richter \etal~\cite{rgb-ilisi-22}   & '21-'22   & 14m        & (CDN logs) & \xmark & \xmark & \xmark & n/a          &         & 2.04B \\
	  Ronan \etal~\cite{rm-rrint-23}      & '22     & 6m         & /48  & \cmark  & \xmark    & \xmark & /48 &        	  & 5.13k \\
	  Zhao \etal~\cite{zkf-edpfi-24}      & '23     & 6m         & /56, 12$\times$/64 & \xmark  & \xmark & \cmark & /48 &  DNS exposure  & 33M \\
    \midrule
	  This work                           & '23-'24     & 11m        & /32, 3$\times$/48& \cmark & \cmark & \cmark &  /29, /32--/48, /48 & Productive subnet 	&   51M\\
    \bottomrule
  \end{tabular}
\end{table*}

\paragraph{Temporary network telescopes}
Czyz \etal~\cite{clmbk-uiibr-13} analyzed IPv6 Internet background radiation~(IBR) by announcing five /12 address blocks  assigned to each of the five RIRs---the covering prefixes of the full address space allocated by LIRs.  
This experiment extracted darknet traffic, \ie packets to the subnets (of the /12) that were never allocated nor routed.
This darknet received only 5\% (209M) of all packets, as most packets were sent to address sub-spaces that was previously allocated by others.
Strowes \etal~\cite{swosf-dawvn-20} analyzed traffic to a newly announced /12 prefix for one week to check visibility and reachability in IPv6 routing.
4$\times$ /32 and 4$\times$ /48 prefixes of a covering /29  were announced separately and actively probed with ICMPv6 packets, which generated most of the received traffic.
In contrast, we  focus on the visibility of BGP~announcements for scanners at network telescopes. %

\paragraph{Server logs}
Richter \etal~\cite{rgb-ilisi-22} analyzed IPv6 scanning activities using the firewall logs of a large CDN but exclude ICMP traffic and TCP packets to ports 80 and 443.
Two sources originated 70\% of all traffic.
75\% of scan sources (aggregated by /64 source prefix) only focused on addresses discoverable via DNS. %
While authors often used /64 source address aggregation, they concede that aggregation is usually case-specific.
Our grouping of scanner activities relies on scan sessions, which showed to be a stable measure under aggregation.

\paragraph{Attracting scanners on the application layer}
Tanveer \etal~\cite{tspn-gduia-23} examined how IPv6 host activities influence the behavior of IPv6 scanners in a previously unused  /56 subnet. 
Each of the six traffic classes was deployed in four randomly selected /64 subnets.
Publicly visible active services (NTP, Tor, DNS zones) attracted significantly more scan traffic than client activities from the telescopes (web crawls and DNS probes).
Most probes targeted low-byte addresses and random IIDs.
Scanners focused on a few subnets or scanned across all 256 subnets.

Zhao \etal~\cite{zkf-edpfi-24} investigated the effect of address exposure via DNS using a previously unused /56 network. They published addresses via four DNS methods (\eg PTR record for random addresses).
Each method was deployed once in a /64 darknet and once in a /64 honeynet.
Associating IPv6 addresses with domains that have IPv4 PTR records attracted over 99.99\% of the scans.
In honeynets, scanners were less focused, targeting exposed addresses and unexposed addresses in the vicinity more equally. %
Low-byte scans comprised only 0.03\% of all scans.

Parallel to our work, Tanveer \etal~\cite{tcmkr-uisdl-25} deployed a proactive network telescope with multiple application layer attractors to evaluate the effectiveness of these features across the network stack, and to infer the strategy of scanners.

In contrast, this work focuses on the network conditions of proactive and reactive IPv6 telescopes and analyzes how scanners react. We do not proactively attract scanners via specific application services but include a prefix in our analysis that hosts an actively operating~subnet. These controlled properties of our four telescopes allow for selectively characterizing scanner behavior as well as the corresponding bias envisioned in the different telescopes.

\section{Measurement Method and Setup}
\label{sec:method}
\begin{figure}[t]
	\centering
	\includegraphics[width=0.85\textwidth]{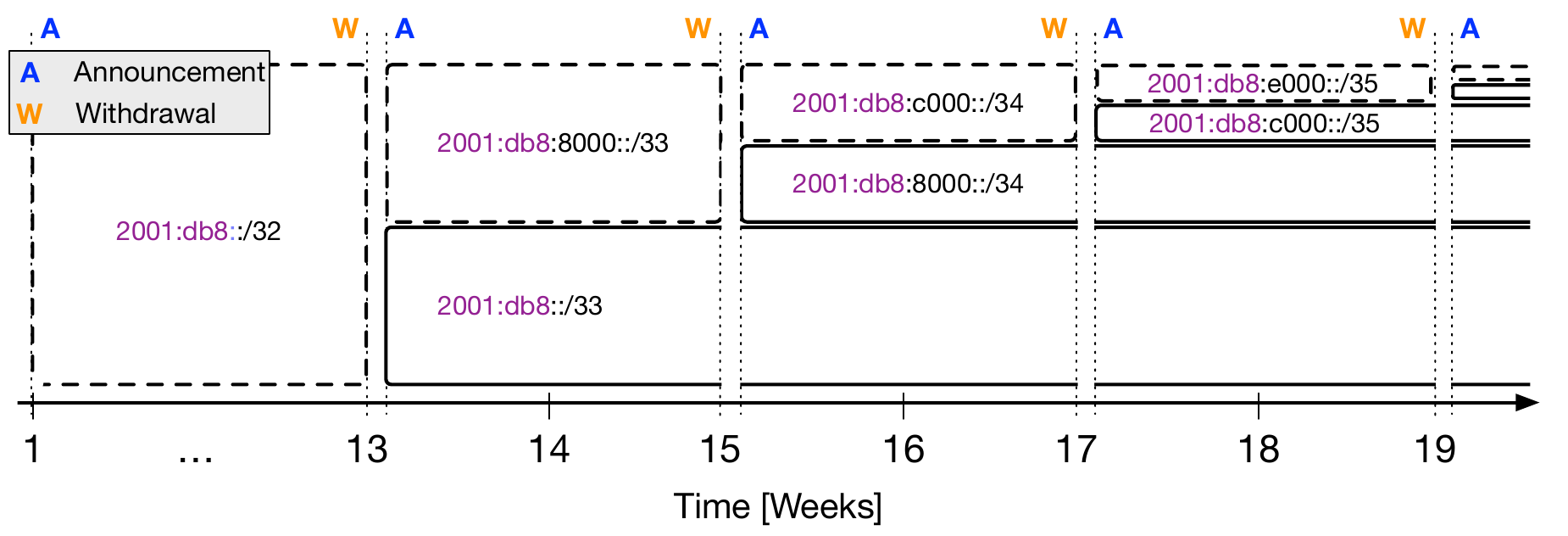}
	\caption{After an initial baseline observation phase (week 1-13), we recursively split one prefix into two more-specific prefixes every two weeks (dotted vertical lines) until we announce 17 prefixes, our most-specific prefix is /48. (\texttt{2001:db8::/32} is not the prefix we announce but reserved for documentation.)}%
	\label{fig:method:splitting}
\end{figure}
\begin{figure}
	\centering
  \resizebox{0.9\textwidth}{!}{%
        \input{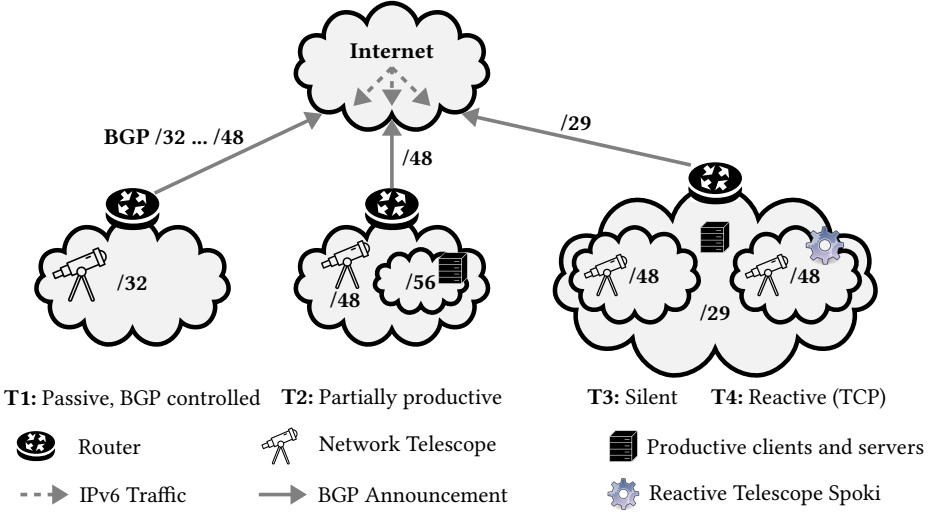}
    }
  \caption{Our experiment setup, showing all four network telescopes, their sizes, and announced prefixes in~BGP.}
	\label{fig:method:telescope-setup}
\end{figure}

In this section, we introduce our measurement infrastructure consisting of four different network telescopes (T1--T4). Each telescope has specific properties that reflect different aspects of network embedding (see \autoref{fig:method:telescope-setup}). We define our network experiments along with our scanner detection method. 

\subsection{Network Telescopes}
Scanning success in the IPv6 address space requires strategy and guidance, reasonably based on external observations and measurements. 
Hitlists and target generation algorithms can help to reduce the search space, but scanners may pursue different strategies, which we want to discover.

\paragraph{T1: BGP controlled /32 -- /48}
This untainted /32 IPv6 prefix was first announced at the beginning of our experiments.
All addresses in this prefix are passive, and neither is assigned to an endpoint nor to other network services such as the DNS. In the first 12~weeks of our experiment, this prefix was kept stable. During this initial observation period, we collected baseline data.
Thereafter, we used this prefix to study the influence of BGP announcements, announced prefix sizes, as well as the number and relative position of (sub-)prefixes on scan activities.
For this active, controlled experiment we carefully selected the duration of the announcements, as well as the number and size(s) of the announced prefix(es), to avoid  interference with scanner behavior in any other way.

\begin{wrapfigure}{r}{0.5\textwidth}
	\centering
	\vspace{-13pt}
	\includegraphics[width=.5\textwidth]{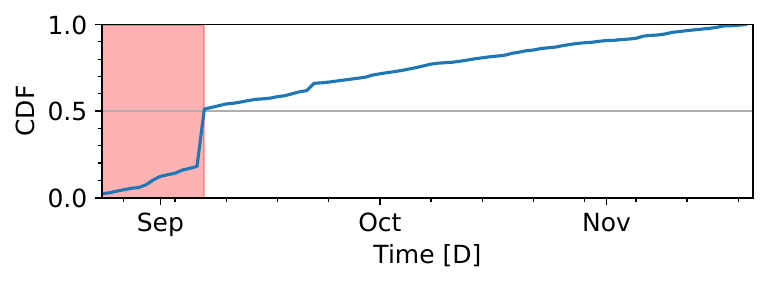}
	\caption{Number of new prefixes hosting scan sources discovered during our initial 12-week observation~period. Two weeks (red area) mark the trade-off to learn enough without running an announcement too long.}
	\label{fig:method:new-prefixes}
\end{wrapfigure}
To identify a reasonable announcement period, we explore the temporal correlation of scanners with a new BGP announcement during our initial observation. We do not find convergence, but after two weeks the number of scan sources from new prefixes reduces notably (see \autoref{fig:method:new-prefixes}). Thereby, we decide on a two-weeks prefix split interval as a trade-off between overall experiment duration and  time to observe how scanners react to our~changes.

Every two weeks, we withdraw our prefixes for one day.
On the day after, we announce a set of new prefixes that contains
\one a new pair of prefixes created by splitting one prefix into two more-specific prefixes of equal size and
\two all previously announced prefixes except the covering prefix. 
Since low-byte addresses tend to attract more traffic, we chose to split the most-specific prefix that does not contain the respective low-byte address if possible (\eg the low-byte address of \texttt{2001:db8::/32} is \texttt{2001:db8::1}).
This creates two new prefixes with low-byte addresses that do not (byte-wise) match the low-byte addresses of the previously announced prefixes.
The number of prefixes we announce increases by one at each interval, including two previously unannounced prefixes.  \autoref{fig:method:splitting} visualizes our asymmetric prefix splitting process.

\paragraph{T2: Partially productive /48}
This /48 prefix has been continuously announced for 13 years and a /56 subnet (not separately announced) is in productive use since then. The productive subnet contains web servers, end hosts, and IoT devices, several of which with persistent DNS entries; traffic from or to the /56 subnet is excluded from our measurements. In addition, one address within the /48 prefix has a DNS entry outside the active /56 sub-prefix. This name co-exists in IPv4 and is part of the CISCO Umbrella popularity list~\cite{cisco_umbrella_top_1m}.

\paragraph{T3: Silent /48}
We borrowed this /48 network, which is not separately announced in BGP, as part of a larger /29 prefix. This network neither hosts services nor active clients but is entirely silent.

\paragraph{T4: Reactive /48}
This /48 is part of the same /29 covering prefix as T3. In contrast to T3, T4 answers to TCP SYN packets. T4 deploys the reactive network telescope Spoki~\cite{hnkds-sunws-22}, which interacts generically on the transport layer. Spoki accepts TCP connections and reacts to SYN scanning for analyzing scanner behavior. First results from Spoki can be found in \cite{xkehs-ptsif-25}.

\subsection{Experiment Setup}

We run Free Range Routing (FRR)~\cite{frrouting} software on Linux  to connect our autonomous system to an IXP and to upstream providers.  BGP announcements for T2---T4 remain stable during our measurement period.
For T1, the bi-weekly (re-)configuration is performed automatically to avoid errors.
We confirm visibility of our announcements via a looking glass~\cite{telia-lg} and RIPEstat~\cite{ripestat}.

\paragraph{Route6 object and ROAs}
Route(6) objects are information records that publish peering relations in the RIR database and are often used in public peering, occasionally also by upstream providers to validate that routes received from their peers are legitimate. To assess the impact of route objects on scanners, we first omitted this for our initial /32 prefix (T1). This did not impair the visibility of our prefix via the upstream.
Four months after its first announcements,  we created a route object for the non-split /33 prefix.
Creating the route object had no noticeable effect on scanners. 
We did not add an RPKI ROA~\cite{RFC-6480}, since this would neither enhance reachability nor visibility of our prefixes given that prefixes validated as \emph{not found} are not filtered. %

\paragraph{Presence in the TUM hitlist}
The TUM hitlist service~\cite{gsgc-siitc-16,zssgc-rcdir-22} is the most popular IPv6 address collection of active hosts.
It comprises lists of responsive addresses, aliased prefixes, and non-aliased prefixes. 
If present on the TUM~hitlist, the prefixes from our telescopes T1 to T4 could see enhanced visibility to scanners.
At the beginning of our experiment, T1 was absent but T2 as well as the covering /29 prefix of T3 and T4 were already listed on the hitlist (non-aliased prefixes). We first observed the /32 prefix of T1 on the same hitlist on August 29, 2023---5 days after its announcement. We report on the presence of new prefixes from T1 on the hitlist in \autoref{sec:splitting}. It is noteworthy that T4---even though responsive from every address---never appeared on the aliased prefix list.

\subsection{Scanner, Sources, and Sessions}
\label{sub:source-aggregation}

We now specify terminology concerning scanners and explain our methods of identification.

\paragraph{Scanner and scan sources}
A scanner may send packets from a single source, or from locally as well as globally distributed addresses. Identifying globally situated scanning entities is a complex task and we leave this for future work. Instead, we focus on localizable scan sources in this~work.

\begin{wrapfigure}{r}{0.48\textwidth}
	\centering
	\includegraphics[width=0.48\textwidth]{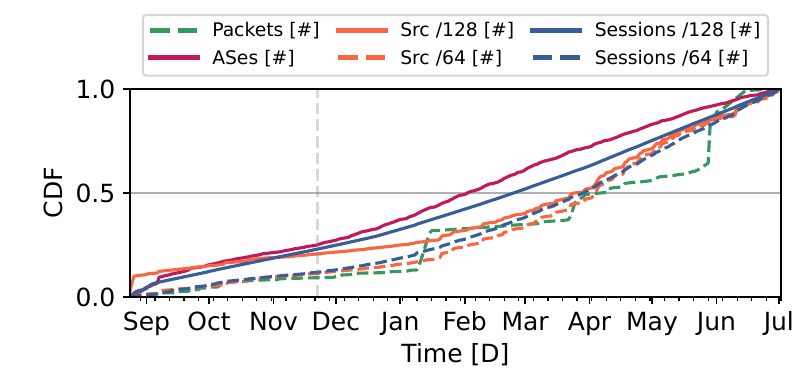}
	\caption{Relative number of packets, ASes, and sources~(/128, /64) learned during our measurements aggregated over all 4 telescopes. The vertical line indicates the end of our initial observation period.}
	\label{fig:method:full-period-all-telescopes}
\end{wrapfigure}
A localizable \emph{scan source} is an individual IPv6~address or an aggregation of addresses from some network. 
Inspecting addresses individually, \ie a /128, is the most fine-grained view.
Aggregating sources based on network addresses (/64) can help to reveal scanners that change addresses within their subnet.
Aside from local subnets, sources may be also aggregated on a prefix level, \eg in a /48~\cite{ripe-labs-visibility-ip-prefixes}.
Aggregating sources in larger prefixes than /48 may likely combine  unrelated scan sources, especially for hosting networks.
Most  related work~\cite{clmbk-uiibr-13,swosf-dawvn-20,lhlbl-intnt-21,rgb-ilisi-22,zkf-edpfi-24,rm-rrint-23} identifies scanners by addresses (/128), followed by aggregated /64 network prefixes~\cite{swosf-dawvn-20,rgb-ilisi-22}. Richter \etal~\cite{rgb-ilisi-22} additionally evaluated /48 scan sources.

We examine the influence of different aggregation levels on our telescopes, see \autoref{fig:method:full-period-all-telescopes}. We find divergence between the aggregation levels /128 and /64. 
For this reason, we proceed by analyzing both, full source addresses (/128) and /64 aggregation. We report in detail on the scan sources and their origin in \autoref{sec:splitting}.

\paragraph{Scan sessions}
We define a \textit{scan session} as a sequence of consecutive packets from a single source, in which the inter-arrival time between two subsequent packets remains below a timeout value \textit{T}. Benson \etal~\cite{bdcsk-libro-15} used a timeout value of 5~minutes for IPv4---the time needed to traverse the entire address space. IPv6 requires an adjustment of this value to cope with scanners traversing randomly through large subnets.  Richter \etal~\cite{rgb-ilisi-22} as well as Zhao \etal~\cite{zkf-edpfi-24} selected one hour as session timeout to trade-off between too loose packet aggregation and too long sessions, which we adopt. %
Nevertheless, we do not apply any other constraints, such as a minimum number of packets or targets per session, since we want to consider all packets of all sources. 

In our subsequent analyses, we will focus on  scan sessions instead of packets, as sessions are more expressive and consolidate heavy hitters. Also, sessions remain relatively stable for the different source aggregation levels (/128 and /64). Similar to the scan sources, the number of scan sessions for full addresses has a more pronounced increase over /64 aggregation (shown in \autoref{fig:method:full-period-all-telescopes}). We report on the correlation between the high number of /128 scan sources and sessions in \autoref{sec:splitting}.

\section{Overview of Data Corpus}
\label{sec:traffic}

We now introduce the data corpus collected during our experiments. After a brief overview of the traffic observed during the initial observation period, we present an aggregated view of the full observation period regarding target address types, protocol, and port usage. We use the free MaxMind GeoIP database~\cite{maxmind-geolitecountry} to geolocate scan sources and the freely available RouteViews dataset~\cite{routeviews} for IPv6 address to ASN mapping.

\subsection{Initial Observation Period}
We captured about 4.6M packets during the first 12 weeks across all telescopes. These packets originate from 7.6k source addresses (3k /64 subnets) and can be aggregated into 168k (/128) or 19k (/64) sessions, respectively.
We observe scan sources in $666$ ASes distributed over 89~different~countries. 

\begin{wrapfigure}{r}{0.48\textwidth}
	\centering
	\includegraphics[width=0.48\textwidth]{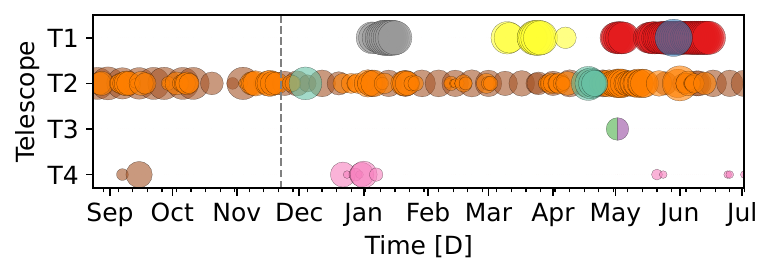}
	\caption{Daily activity from heavy hitters at the four  telescopes. Bubble sizes signify packet counts and colors represent scan sources (same color means same scan source).}
	\label{fig:heavy-hitter-per-day-bubbles}
\end{wrapfigure}
\subsection{Full Observation Period}
In total, we captured over 51M packets across all telescopes, originating from 36k  source addresses (26k /64 subnets). We can assign packets to 754k (/128) and 151k (/64) sessions, respectively.
Scan sources originate from 2k ASes in 127 different countries.
All CDFs grow smoothly over the whole time except for the number of packets, which discontinuously grows due to some heavy hitters sending large amounts of packets within short periods of time, see \autoref{fig:method:full-period-all-telescopes} and \autoref{sec:behavior} for details. %

\paragraph{Heavy hitters}
A few individual  sources stand out by contributing more than 10\% of the scan packets at one telescope. We found ten such heavy hitters  at the four telescopes, four in T1, three in T2, two in T3, and two in T4 (one source is identified as a heavy hitter in T2 and T4). Most heavy hitters send large amounts of packets in very few sessions during a short time span. Two heavy hitters in T2, however, send packets repeatedly over the complete observation period (see \autoref{fig:heavy-hitter-per-day-bubbles}), one of which has an RDNS entry pointing to the 6Sense scanning campaign~\cite{wehbb-6iiss-24}.
Only three heavy hitters have an RDNS entry but we can attribute seven out of ten heavy hitters to a research context. Three of four heavy hitters in T1 are located in networks of hosting providers, one of them  claims to be a `bullet-proof' hosting provider, indicating that its scan campaign  might be of malicious intent.
We do not exclude heavy hitters from our analysis as they do not dominate our session-centric statistics.
Even though heavy hitters account for 73\% of all packets, they only contribute 0.04\% of all sessions.

\paragraph{Target address types}
To gain a better understanding of the selected destination addresses, we categorize addresses using the \textit{addr6} tool of the \textit{IPv6Toolkit}~\cite{sn-it-16}. 
The address types are defined
\begin{wraptable}{r}{0.5\textwidth}
	\centering
	\footnotesize
	\vfill
	\caption{Packets, sessions, and sources per transport protocol. Shares of sessions and sources may exceed 100\% due to multi-protocol scanners.}\vspace{-10pt}
	\label{tbl:traffic:proto-usage}
	\begin{tabular}{lrrrrrr}
		\toprule
		& \multicolumn{2}{c}{Packets} & \multicolumn{2}{c}{Sessions /128} & \multicolumn{2}{c}{Sources /128} \\
		\cmidrule(lr){2-3}
		\cmidrule(lr){4-5}
		\cmidrule(lr){6-7}
		Protocol & \makecell[c]{[\#]} & \makecell[c]{[\%]} & \makecell[c]{[\#]} & \makecell[c]{[\%]} & \makecell[c]{[\#]} & \makecell[c]{[\%]}
		\\
		\midrule
		ICMPv6 & \num{33889898} & 66.2 & \num{132816} & 20.1 & \num{20373} & 56.5 \\
		UDP & \num{11967255} & 23.4 & \num{36780} & 5.6 & \num{7113} & 19.7 \\
		TCP & \num{5372494} & 10.5 & \num{614223} & 92.8 & \num{19977} & 55.4 \\
		\bottomrule
	\end{tabular}
\end{wraptable} 
\noindent
according to the specifications of RFC~7707~\cite{RFC-7707}. Additionally, we identify destination addresses ending with \texttt{::0} as Subnet-Router anycast addresses according to RFC~4291~\cite{RFC-4291}. We observe most targeted addresses to be randomized ($64\%$), followed by low-byte ($23\%$), and pattern-byte addresses~($6\%$), see \autoref{tbl:addrtypes-target}. While randomized addresses comprise most packets, they only account for 6\% of the scanners. 90\% of all scanners target at least one low-byte address, showing a clear trend in the scanning strategy. We analyze these results in detail in Sections \ref{sec:behavior} and \ref{sec:splitting}.

\paragraph{Protocol usage}
We group the  packets per transport protocol. Most scanners use ICMPv6 as the primary protocol.
We observe 33.9M ICMPv6 packets, 12M UDP~packets, and 5.4M TCP packets.
$85\%$ of the UDP packets are DNS requests from a single scanner.
While only $10.5\%$ of the total packets are TCP, they originate from $55.4\%$ of the scan sources and $92.8\%$ of the sessions, implying that TCP is scanned quite often but with small amounts of packets (see \autoref{tbl:traffic:proto-usage}).
In contrast, UDP is probed in only $5.6\%$ of the sessions but accounts for $23.4\%$ of the packets, which shows the impact of heavy hitters on the telescope traffic.

\begin{table}
	\begin{minipage}[t]{0.48\linewidth}
		\centering
		\footnotesize
		\vfill
		\caption{Distribution of target types. Shares may exceed 100\% due to probing multiple address types.}
		\label{tbl:addrtypes-target}
			\begin{tabular}{lrrrr}
				\toprule
				& \multicolumn{2}{c}{Packets} & \multicolumn{2}{c}{Sources /128} \\
				\cmidrule(lr){2-3}
				\cmidrule(lr){4-5}
				Address Type & \makecell[c]{[\#]} & \makecell[c]{[\%]} & \makecell[c]{[\#]} & \makecell[c]{[\%]} \\
				\midrule
				randomized & \num{32911060} & 64.24 & 2101 & 5.83 \\
				low-byte & \num{11828733} & 23.09 & 32350 & 89.71 \\
				pattern-bytes & \num{3054847} & 5.96 & 570 & 1.58 \\
				embedded-ipv4 & \num{2026762} & 3.96 & 547 & 1.52 \\
				subnet-anycast & \num{1173137} & 2.29 & 1476 & 4.09 \\
				embedded-port & \num{138656} & 0.27 & 80 & 0.22 \\
				ieee-derived & \num{96627} & 0.19 & 26 & 0.07 \\
				isatap & \num{217} & $\leq 0.01$ & 2 & $\leq 0.01$ \\
				\bottomrule
			\end{tabular}
	\end{minipage}
	\hfill
	\begin{minipage}[t]{0.48\linewidth}
		\centering
		\footnotesize
		\vfill
		\caption{Top 5 ports targeted by sessions on /64 source aggregation level. Shares add up to >100\% as multiple ports may be addressed in one session.}
		\label{tbl:traffic:port-usage}
		\begin{tabular}{lrrrrrr}
			\toprule
			& \multicolumn{3}{c}{TCP} & \multicolumn{3}{c}{UDP} \\
			\cmidrule(lr){2-4}
			\cmidrule(lr){5-7}
			Rank & \makecell[c]{Port} & \makecell[c]{[\#]} & \makecell[c]{[\%]} & \makecell[c]{Port} & \makecell[c]{[\#]} & \makecell[c]{[\%]} \\
			\midrule
			\#1 & 80    & \num{48070} & 87.2 & Traceroute{\footnotesize$^1$} & \num{7067}  &  71.4 \\
			\#2 & 443   & \num{16223}   & 29.4  &     53     & \num{1945}    & 19.7  \\
			\#3 & 21    & \num{2592}   & 4.7  &     161    & \num{1718}    & 17.4  \\
			\#4 & 8080    & \num{2172}   & 3.9  &     500    & \num{1710}    & 17.3  \\
			\#5 & 22    & \num{1849}   & 3.4  &     123   & \num{1669}    & 16.9  \\
			\bottomrule
			\multicolumn{7}{l}{\footnotesize$^1$Default port range of traceroute [33434, 33523].}\\
		\end{tabular}
	\end{minipage}
\end{table}

\paragraph{Target ports}
We count all targeted destination ports once per session in which they occur. We show the top 5 ports for TCP and UDP packets in \autoref{tbl:traffic:port-usage}.
Since some scanners rotate their source interface IDs per destination port during vertical scans, we aggregate sessions by /64 subnets for this analysis.
Most sessions include TCP packets with destination port 80 (HTTP, 87\%) and 443 (HTTPS, 29\%). The remaining top 5 ports received comparatively little attention~(3\%-5\%). In total, {1,335} TCP ports were hit at least once. Additionally, each of the top 72 ports were targeted in at least 1k sessions, which shows that apart from port 80 and 443 some scanners cover a broad port range during their scans. A different picture emerges for UDP. 71\% of all UDP sessions targeted a port from the standard traceroute range. The remaining top 5 ports belong to DNS, SNMP, ISAKMP, and NTP, which are observed in a similar number of sessions. In total, \num{91} UDP ports (aggregating all traceroute ports) were visited at least once.

\paragraph{Key observations} {Scan traffic arrived continuously in our telescopes during the eleven months of experiments. ICMPv6 is the dominating protocol. Low-byte address scanning was the most popular strategy, but the majority of packets targeted randomized addresses. HTTP ports (TCP 80 and 443) are probed frequently while UDP packets predominantly target common traceroute and DNS ports.}

\section{Taxonomy of Scanning Behavior and Public Scan Tools}
\label{sec:taxonomy}

Next we introduce our classification of IPv6~scanners.
We categorize \textit{temporal behavior} of scanners, \textit{network selection}, \ie properties of target networks, and \textit{addresses selection}, \ie properties of the IPv6 target interface IDs.
In addition, we classify IPv6 scan tools based on their signatures left in payloads and RDNS entries.
These classifications are applied to scan sessions (\cf \autoref{sub:source-aggregation}).

\subsection{Classifying Temporal Behavior}
\label{sub:taxonomy:temporal}

\begin{figure}[t]
	\centering
	\begin{tikzpicture}[x=0.75cm,y=0.75cm]
	
	\draw[->,line width=1.5pt] (0,0) -- (4,0) node[midway,below=0.1cm] {One-off} node[right,left=3.9] {$t$};
	\foreach \x in {1,2,...,3} {
		\draw (\x,-0.1) -- (\x,0.1);
	}
	
	\draw[->,line width=1.5pt] (6,0) -- (10,0) node[midway,below=0.1cm] {Periodic} node[right,left=3.9] {$t$};
	\foreach \x in {7,8,...,9} {
		\draw (\x,-0.1) -- (\x,0.1);
	}
	
	\draw[->,line width=1.5pt] (12,0) -- (16,0) node[midway,below=0.1cm] {Intermittent} node[right,left=3.9] {$t$};
	\foreach \x in {13,14,...,15} {
		\draw (\x,-0.1) -- (\x,0.1);
	}

	\draw[fill=black] (1,0.4) circle (0.1);
	
	\foreach \x in {7,8,9} {
		\draw[fill=black] (\x,0.4) circle (0.1);
	}
	
	\foreach \x in {12.5,13.3,15.1} {
		\draw[fill=black] (\x,0.4) circle (0.1);
	}
	
\end{tikzpicture}
	\caption{Classification of temporal scanner behavior.}
	\label{fig:taxonomy:temporal_classification}
	\vspace{-10pt}
\end{figure}
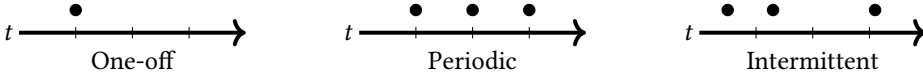

The temporal behavior may get influenced by internal schedules and external events.
We classify scanners into three basic categories as visualized in \autoref{fig:taxonomy:temporal_classification}.

\begin{description}

\item[One-off]
 scanners perform a single scan session and then disappear for the remaining measurement.
Two edge cases are captured in this category: \one scanners that perform slow, long-lasting scans and thus maintain a single scan session and \two scanners that alter source addresses repeatedly during a session, provided we can attribute the individual sessions.

\item[Periodic]
 scanners perform scan sessions in periodic intervals.
The period can vary between scanners from hours to months but scanners must appear more than twice and show a stable period between scans.

\item[Intermittent]
 scanners perform multiple scan sessions but do not show a periodic pattern.
Specifically, such scanners must appear with at least two scan sessions in our dataset but show no detectable period between scans.

\end{description}

\paragraph{Classification method}
Scanners fall into exactly one of the three exclusive categories.
We identify  \textit{periodic} scanning  based on a period detection by autocorrelation~\cite{bwrj-omdpr-23}, which leaves those recurrent scanners as \textit{intermittent} for which no period is~detectable.

\subsection{Classifying Network Selection}

Scanners implement a strategy for probing IPv6 network ranges, which they can reach as soon as their prefixes are announced in BGP.
We are interested in the target selection of network addresses by scanners with a particular focus on scanner behavior related to BGP prefix announcements.
A scanner follows either a single-prefix or a multiple-prefix scanning strategy, and then scans the selected networks independent or dependent of their size or in a mix of both. In our setup, we can only distinguish between network-size independent and dependent scanning for multiple-prefix scanners as our view point is the context of arriving probe packets.

\begin{description}
\item[Single-prefix scanning]
A scanner that only scans one announced pre\-fix during each period of announcement in BGP.
The chosen (arbitrary) prefix may vary between periods.

\item[Network-size independent]
A scanner hits networks of different sizes with (roughly) the same number of scan sessions.
This behavior becomes particularly visible in our BGP experiments (T1) since all but two of the prefixes differ in size by a factor of at least two.
For example, a scanner that continuously traverses the  entire address space belongs in this category as each prefix will see the same number of sessions.
Since our classification is based on sessions, it does not matter how many packets are sent into each prefix.

\item[Network-size dependent]
A scanner varies the number of sessions based on the network size.
In this category, we expect scanners to target smaller networks and more-specific prefixes with significantly fewer scan sessions as they contain fewer addresses.
Examples are course-grained scans that more likely hit less-specific prefixes. 
Also, scanners may direct sessions toward fixed-size subnets. 
This allows scanners to probe a specific share of addresses per network and  add sessions to increase the coverage of less-specific prefixes.

\item[Inconsistent]
A scanner that changes its behavior during announcement periods shows inconsistent behavior and does not clearly fit into any of the categories above.
\end{description}

\paragraph{Classification method}
For T2--T4, a scanner can only be classified as single-prefix scanner.
For~T1, we determine the network selection strategy separately for each announcement cycle, \ie the two weeks during which we announce a specific set of prefixes in T1.
Our classification is based on the density-based clustering algorithm~DBSCAN.

\subsection{Classifying Address Selection}

For a given network prefix, a scanner needs to select endpoint addresses (notably interface IDs) for probing. 
We categorize the target selection strategy of a scanner per scan session in three groups.

\begin{description}
\item[Structured] 
A scanner selects addresses with a detectable pattern (or strong tendency) towards a specific address structure.
This includes a focus on low-byte addresses, \ie choosing to probe the \texttt{::x} addresses of announced prefixes, following other well-known IPv6 address structures (\cf \autoref{sec:background}), or algorithmically subdividing and iterating prefixes.

\item[Random] We examine the randomness of selected addresses in scan sessions using a statistical test suite, which differs from previous studies by Richter \etal~\cite{rgb-ilisi-22}. 

\item[Unknown]
If no detectable pattern is visible, we classify address selections as unknown.
\end{description}

\paragraph{Classification method}
We test for address structures using the \textit{addr6} tool of the \textit{IPv6Toolkit}~\cite{sn-it-16} and for randomness using the frequency test from the NIST~test~suite~\cite{brsns-stsrp-10}.
For statistical testing, we select sessions of at least 100 packets from our dataset, since the test requires a minimum of bits as input.
The test calculates a p-value between 0~(non-random) and 1 (random).
We consider scans with a p-value of at least 0.01 to use randomly generated addresses, \ie significance level of $\alpha = 0.01$. 
If target sets neither show structure, nor randomness we classify it as unknown.

\subsection{Scan Tools and Systems}

Probes sent by scan tools can carry tool-specific payloads. We cluster the hex-byte representation using DBSCAN, a density-based clustering approach.
Then, we analyze the payload features of each cluster manually by matching them against common publicly measurement tools and systems (\eg \textit{RIPE Atlas probes}~\cite{ripe-atlas}, \textit{Yarrp}~\cite{b-yirhs-16}, and \textit{traceroute}). 
To consider less common tools, we also search \textit{GitHub} repositories and publications on measurement tools. In addition to payload analysis, we execute reverse DNS~queries for each scan source to gain more detailed information on who is scanning our telescopes. We label each cluster according to our tool and DNS analysis. If no reverse DNS~entry exists and no public tool is found, we categorize them using other payload characteristics \eg random bytes, or by  additional patterns in the scan behavior \eg address~rotation. 

\section{Scanners at the Network Telescopes During the Initial Observation Period}
\label{sec:behavior}

\begin{table}
	\caption{Comparison of network telescopes before the split period.}
	\begin{subtable}{0.48\linewidth}
		\centering
		\footnotesize
		\caption{Sources, ASes, destination addresses, and total number of packets. T2 receives 14\% more packets than T1, and is probed by 380\% more /128 sources.}
		\label{tbl:behavior:total-numbers}
		\begin{tabular}{lrrrr}
			\toprule
			\multicolumn{1}{c}{} & \multicolumn{1}{c}{T1} & \multicolumn{1}{c}{T2} & \multicolumn{1}{c}{T3} & \multicolumn{1}{c}{T4}\\
			\midrule
			/128 Source addr.  &  \num{1386}   &   \num{6611}         & \num{7}  &   \num{253}  \\
			/64 Source addr.   &  \num{1199}   &   \num{2113}         & \num{6}  &   \num{251}  \\
			ASN                &  \num{418}    &   \num{478}          & \num{6}  &   \num{9}    \\
			Destination addr.  &  \num{796443} &   \num{714169}       & \num{20} &   \num{1817}  \\
			Packets				& \num{2161354} & \num{2464417} & \num{43} & \num{3416} \\
			\bottomrule
		\end{tabular}
	\end{subtable}
	\hfill
	\begin{subtable}{0.48\linewidth}
		\centering
		\footnotesize
		\caption{Distinct sources per transport protocol. Percentages add to more than 100\% as a single source can probe multiple protocols.}
		\label{tbl:behavior:src-per-protocol}
		\begin{tabular}{lrrrrrrrr}
			\toprule
			& \multicolumn{2}{c}{T1} & \multicolumn{2}{c}{T2} & \multicolumn{2}{c}{T3} & \multicolumn{2}{c}{T4} \\
			\cmidrule(lr){2-3}
			\cmidrule(lr){4-5}
			\cmidrule(lr){6-7}
			\cmidrule(lr){8-9}
			\multicolumn{1}{c}{Protocol} & \makecell[c]{[\#]} & \makecell[c]{[\%]} &  \makecell[c]{[\#]} & \makecell[c]{[\%]} & \makecell[c]{[\#]} & \makecell[c]{[\%]} &  \makecell[c]{[\#]} & \makecell[c]{[\%]} \\
			\midrule
			ICMPv6   &  \num{1110} & 80.1 &   \num{4112} & 62.2      & \num{7} &   100  & \num{246} & 97.2 \\
			TCP      &  \num{40} & 2.9    &   \num{5311} & 80.3     & \num{0} &  0     &  \num{6} & 2.4 \\
			UDP      &  \num{266} & 19.2  &   \num{1768}  & 26.7    & \num{0} &  0     & \num{1} & 0.4  \\
			\bottomrule
		\end{tabular}
	\end{subtable}
\end{table}

In this section, we analyze the network traffic that each of the four telescopes attracts during the initial observation period.
We will present the results during our BGP experiment later in \autoref{sec:splitting}.

\begin{figure}
	\centering
	\begin{subfigure}{0.48\textwidth}
		\centering
		\includegraphics[width=1\textwidth]{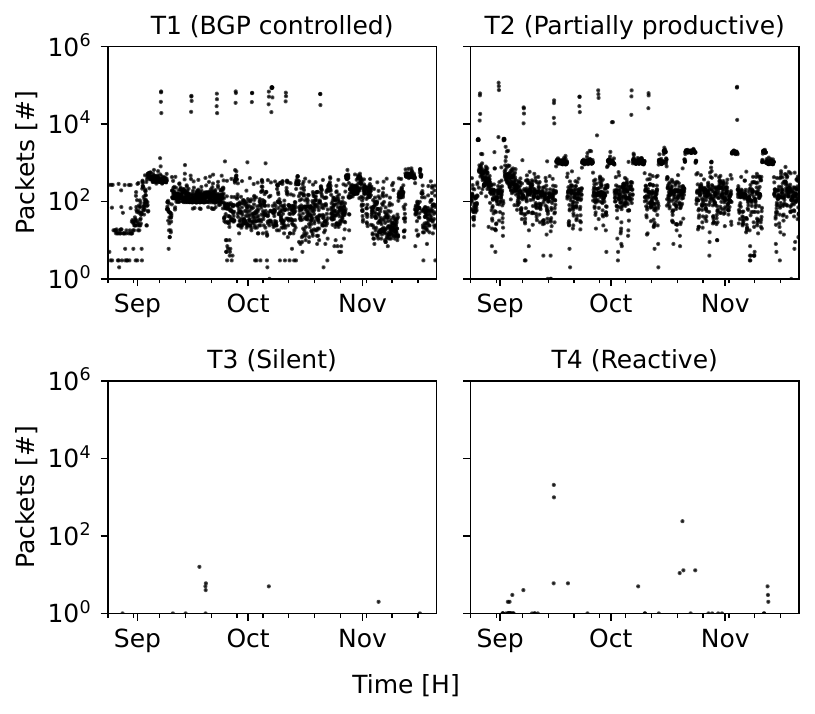}
		\caption{Network traffic per hour across the four telescopes.}
		\label{fig:method:traffic-of-all-telescopes}
	\end{subfigure}
	\hfill
	\begin{subfigure}{0.48\textwidth}
		\centering
		\includegraphics[width=1\textwidth]{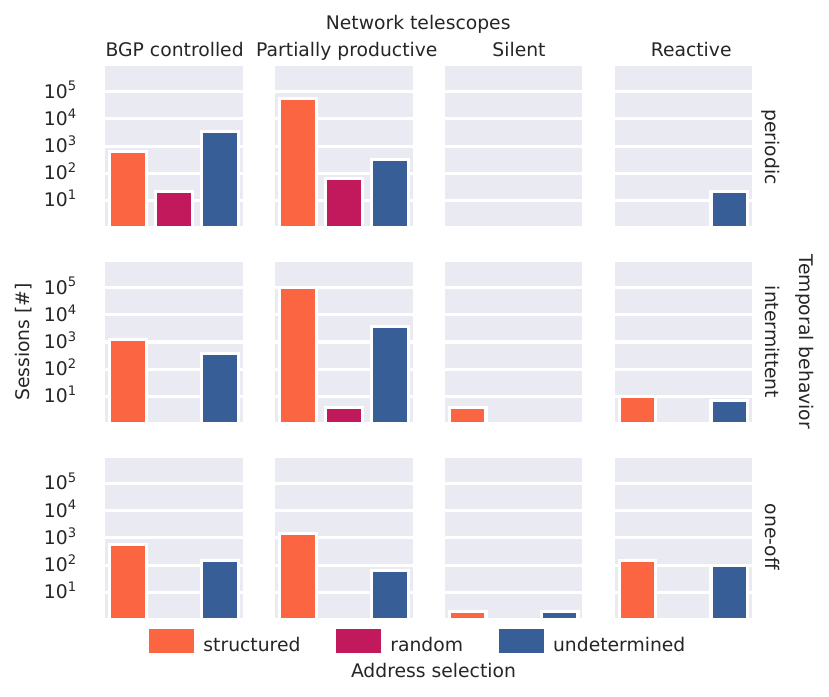}
		\caption{Classification of scanners according to our taxonomy proposed in \autoref{sec:taxonomy}.}
	\label{fig:behavior:taxonomy-before-split}
	\end{subfigure}
	 \vspace{-3pt}
	\caption{Scan traffic and scanning behavior during the initial 12~weeks.}
	\vspace{-12pt}
\end{figure}

\paragraph{Network traffic}
We notice  in \autoref{tbl:behavior:total-numbers} that the two network telescopes with prefixes separately announced in BGP (T1 and T2) receive 4 to 6 orders of magnitude more traffic than those embedded in a covering  BGP announcement (T3 and T4). The reactive network telescope T4, on the other hand, receives two orders of magnitude more traffic than the completely silent T3. 
These observations motivate questions about how BGP announcements impact scan traffic.

\autoref{fig:method:traffic-of-all-telescopes} depicts longer and higher traffic peaks for T2, which emerge from scanners specifically targeting the only address in T2 for which a DNS entry exists.
50\% of all observed scanners exclusively target this address, leading to twice as many  /64 sources visible in T2 than in T1. 
In parallel, T2 scanners exhibit a unique characteristic in source address utilization. \autoref{tbl:behavior:total-numbers} shows little difference in the two source aggregation levels (/64, /128) for T1, T3, and T4, whereas  T2 sees three times as many  /128 scan sources as /64, because T2 attracts scanners that use address rotation within a /64 subnet.  
 The partially productive telescope (T2) receives 14\% more packets and is probed by 380\% more individual sources than our BGP controlled  T1. In contrast, scanners probe 12\% more distinct targets in T1 compared to T2. 
Furthermore, the majority of individual scan sources probes ICMPv6 for T1 (80\%), T3 (100\%), and T4 (97\%). For T2, however, the prevalent probed protocol is TCP (80\%), followed by ICMPv6 (62\%), as shown in \autoref{tbl:behavior:src-per-protocol}. The differences are mainly caused by the DNS attractor and scanners rotating their source addresses within a /64 subnet, which we exclusively observe for T2.

\begin{wrapfigure}{r}{0.48\textwidth}
	\vspace{-12pt}
	\centering
	\includegraphics[width=0.48\textwidth]{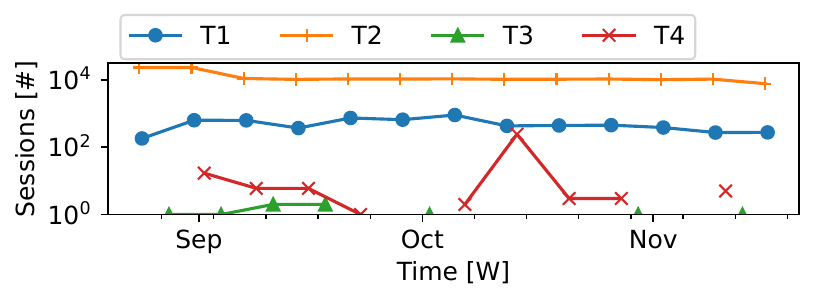}
	\caption{Weekly scan sessions at the four telescopes.}
	\label{fig:sessions-telescopes-12week}
\end{wrapfigure}

\paragraph{Behavior of scanners} \autoref{fig:sessions-telescopes-12week} displays the evolution of session intensities per telescope and week. Results are rather stable for T1 and T2, while sporadic for T3 and T4. The single October peak for T4 is due to a single scanning campaign.   

We next categorize scanners and their sessions according to our taxonomy in \autoref{sec:taxonomy} w.r.t. \one~temporal behavior and \two address selection. \autoref{fig:behavior:taxonomy-before-split} summarizes the classification for all telescopes.
For each telescope (columns), we separate sessions by the temporal behavior of the scanner (rows), and plot bars representing the number of sessions per address selection strategy.
Most scanners return  (intermittent: 41\% or periodic: 29\%) and follow a \textit{structured} selection strategy. 
T1 and T2 appear similar in their overall distribution.
In contrast to T4, scan sessions from \textit{one-off} scanners are relatively less frequent in T1 and T2 where \textit{intermittent} and \textit{periodic} are more common.
For T3 and T4, which have fewer sessions overall, \textit{structured} address selection is the only identifiable strategy. For T3 and T4, none of the sessions are classified as~\textit{random}. 
\begin{figure}
	\centering
	\begin{subfigure}{0.48\textwidth}
		\centering
		\includegraphics[width=1\textwidth]{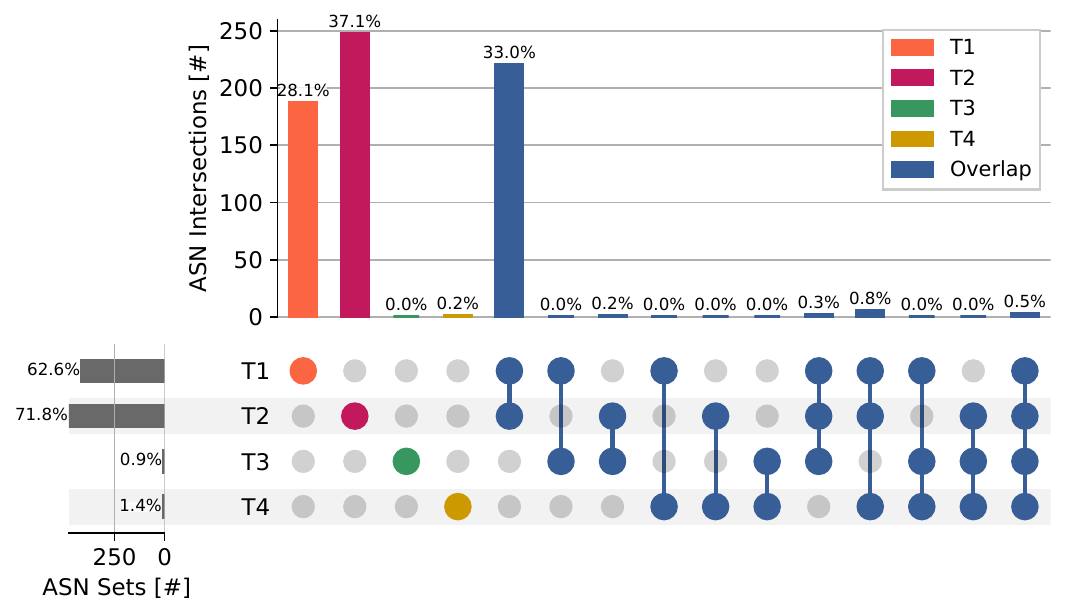}
		\caption{ASN}
		\label{subfig:overlapASN1}
	\end{subfigure}
	\hfill
	\begin{subfigure}{0.48\textwidth}
		\centering
		\includegraphics[width=1\textwidth]{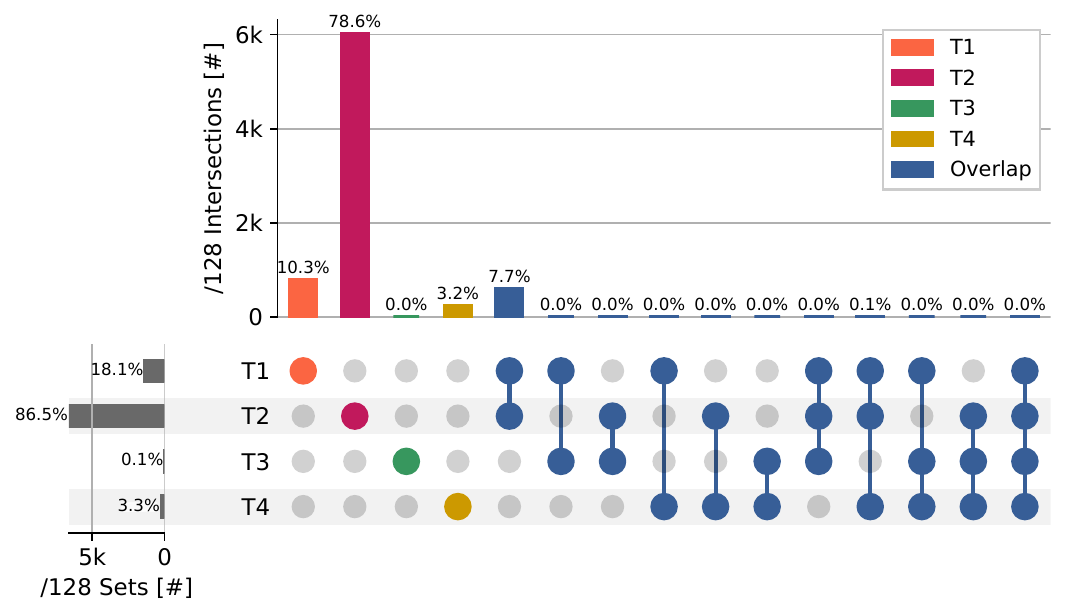}
		\caption{/128}
		\label{subfig:overlap1281}
	\end{subfigure}
	\caption{Intersections of originating autonomous systems and scan sources (/128) between all telescopes.}
\end{figure}

\paragraph{Correlation between the four telescopes}
The initial observation period allows for a brief comparison between all telescopes before exploring the effect of more specific BGP announcements in the following.
At T4, only a small fraction of ASNs is exclusively observed, most of them overlapping with T1-T3.
At T3, all source ASNs can also be observed in T1, T2, and T4.
Around half of all ASNs in T1 and T2 overlap, which shows a partial similarity between these telescopes.
\autoref{subfig:overlapASN1} lays out detailed ASN observations in an UpSet plot.
The bar graph on the left shows the non-exclusive shares observed in each telescope.
In contrast, the top bars show the exclusive combinations of telescopes that observed a given share of ASNs.

A detailed analysis of the /128 scan sources, presented in \autoref{subfig:overlap1281}, reveals that a dominant fraction ($\approx$ 90\%) of the sources exclusively scans a single telescope.
We infer that these differently configured telescopes attract different scanners. 
Even if we exclude all sources that exclusively target the address with a DNS name, diverse traffic remains at all telescopes.

\paragraph{Key observations}
Telescopes that are subnets of covering prefixes in BGP receive significantly less traffic than telescopes that coincide with in BGP announced prefixes. Sessions are a stable measure of scanner behavior, but behavior appears to vary with the kind of scanner attractors.
Most scanners follow a structured address selection strategy and return. Heavy hitters are either research scanners or~malicious.

\section{Adaption of Scanners to BGP Signals}
\label{sec:splitting}
In this section, we report on the detailed results of our BGP experiment. We discuss how traffic is attracted by prefix advertisements and how scanners react to BGP signals.  
We also compare the activities in our BGP controlled telescope (T1)  against those in the other telescopes.

\subsection{BGP-controlled Telescope Activity and Impact on the Behavior of Scanners}
\begin{figure}[t]
	\centering
	\includegraphics[width=\columnwidth]{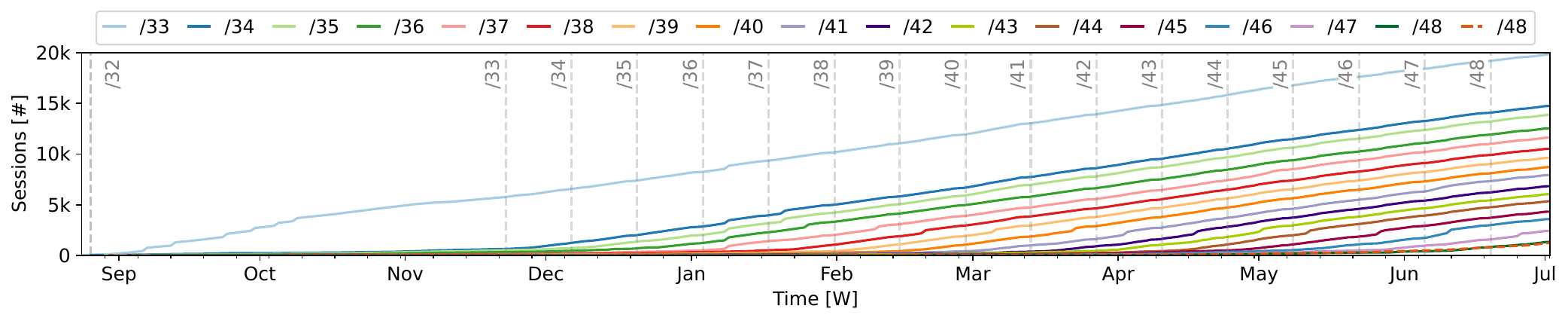}
	\caption{Cumulative number of scan sessions per target prefix. Addresses receive significantly more attention by scanners when more specific prefixes are announced in BGP.}
	\label{fig:splitting:overview-scan-sessions}
\end{figure}

\begin{wrapfigure}{r}{0.5\textwidth}
	\centering
	\includegraphics[width=0.5\textwidth]{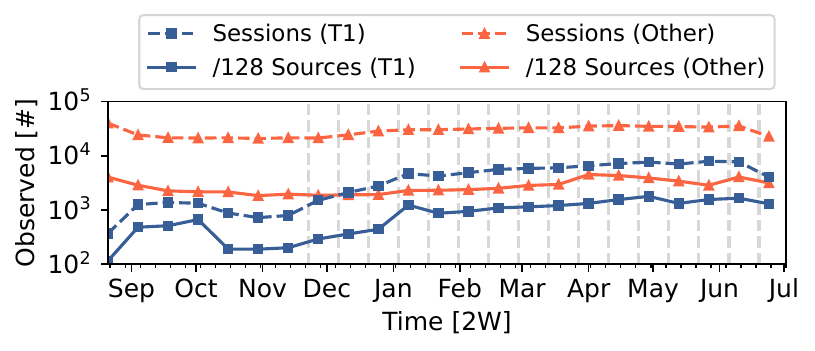}
	\caption{Bi-weekly evolution of sessions and sources in the BGP controlled (T1) and the aggregated remaining telescopes. Different from the other telescopes, T1 shows an increase in sessions and sources during the experiment. The vertical lines indicate the prefix splits.}
	\label{fig:src-sessions-telescopes}
\end{wrapfigure}

\paragraph{Attracting traffic}
During our BGP experiment, T1 observed a weekly increase in the average number of observed scan sources by 275\% and  555\% in the average number of scan sessions. 
Activities remain stable in all other telescopes (see \autoref{fig:src-sessions-telescopes}).
Announcements of the two new more-specific prefixes attract significantly more traffic than to the single announcement of the companion prefix.
\autoref{fig:splitting:overview-scan-sessions} illustrates the growth in sessions per most-specific prefix. As long as subnets remain silent within a larger covering prefix, they hardly attract any attention, \eg  /48 subnets receive only 0.4\% of the total number of sessions during the first two weeks of the experiment. Only after subnets transform to prefixes, they receive significant traffic. The overall number of packets arriving in the iteratively split /33 segment exceeds packet counts for the stable companion /33 by +286\%.
For the final announcement period, the /48 prefixes observed 15.7\% of all sessions (\ie  increased by~39$\times$).

\begin{figure}
	\begin{minipage}{0.45\textwidth}
		\setlength{\tabcolsep}{2pt}
		\centering
		\footnotesize
		\captionof{table}{Classification according to our taxonomy in \autoref{sec:taxonomy}, combining internal and external schedules during the split period.}
		\label{tbl:behavior:source_classification}
		\begin{tabular}{lrrrr}
			\toprule
			& \multicolumn{2}{c}{Scanners} & \multicolumn{2}{c}{Sessions}
			\\
			\cmidrule(lr){2-3}
			\cmidrule(lr){4-5}
			Classification & \makecell[c]{[\#]} & \makecell[c]{[\%]} & \makecell[c]{[\#]} & \makecell[c]{[\%]}
			\\
			\midrule
			\emph{Temporal behavior}
			\\
			\ \  One-off & \num{8244} & 69.71 & \num{8244} & 8.95
			\\
			\ \  Intermittent & \num{1832} & 15.49 & \ \ \num{16842} & 18.28
			\\
			\ \  Periodic & \num{1750} & 14.80 & \num{67067} & 72.78
			\\
			\midrule
			\emph{Network selection}
			\\
			\ \  Single-prefix scanning & \num{10703} & 90.50 & \num{17939} & 19.47  \\
			\ \  Network-size independent & \num{1035} & 8.75 & \num{28433} & 30.85 \\
			\ \  Inconsistent behavior & \num{64} & 0.55 & \num{44294} & 48.07 \\
			\ \  Network-size dependent & \num{24} & 0.20 & \num{1487} & 1.61 \\
			\bottomrule
		\end{tabular}
	\end{minipage}
	\hfill
	\begin{minipage}{0.45\textwidth}
		\centering
		\includegraphics[width=\textwidth]{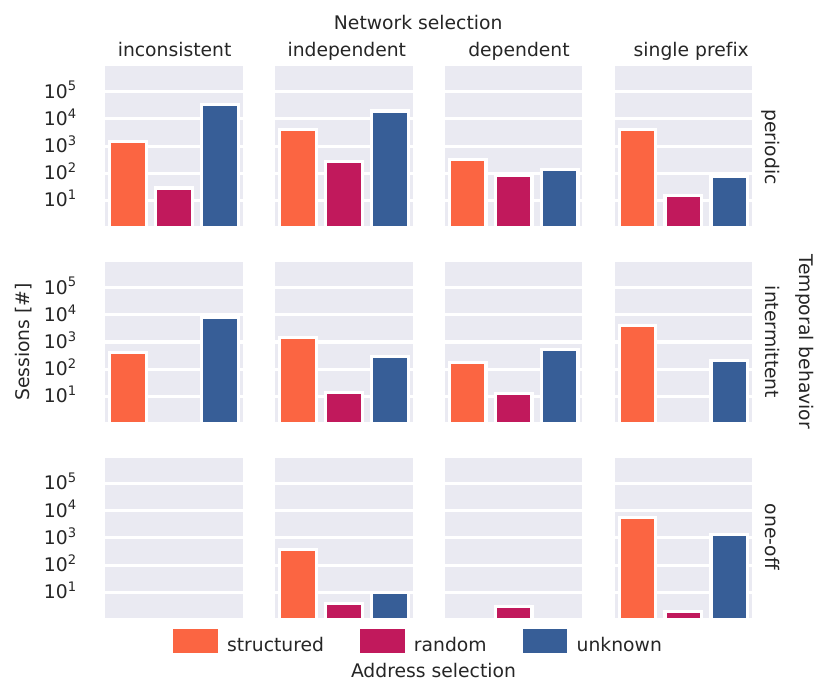}
		\caption{Classification of scanners observed in T1 during the split period according to our taxonomy.}
		\label{fig:behavior:taxonomy_t1}
	\end{minipage}
\end{figure}

\paragraph{Network selection of scanners}
We observe from Table~\ref{tbl:behavior:source_classification}  that 90\% of all scan sources probe addresses from a single prefix, 9\% scan independent of network size, less than one percent show inconsistent or network-size dependent behavior.
Single-prefix scanners originate mainly from two entities \one RIPE Atlas (57\%) topology measurements, and \two Alpha Strike Labs~\cite{alphastrike-labs} (36\%), a cybersecurity company. These scanners account for 20\% of the sessions. 

With more than one thousand individual scan sources and 31\% of the total sessions, a large portion selects target networks independent of the actual prefix size, \ie during an announcement period they cover each of our prefixes with a roughly equal number of scan sessions.
Since the sizes of our BGP prefixes differ largely these scanners either aim at targeting every prefix during each session or they scan the BGP space with fine granularity, hitting our prefixes as a side effect of their strategy. We observe inconsistent behavior for only 64 scan sources, but these execute almost 50\% of all sessions. In most cases, inconsistent behavior tends to become network-size independent towards the end of the experiment. In the beginning, however, these scanners probe the larger prefixes more often.

Only 24 scan sources (0.2\%) probe networks depending on their size, \ie larger prefixes receive more traffic. A minimal telescope setup with a single /48 announcement would miss those who exclusively probe larger prefixes, such as /32.
In contrast, we expect scanners that select their target networks independent of size to be observable in all network telescopes with a prefix visible in BGP. Even smaller telescopes should receive a notable amount of their scan traffic.

\begin{figure}
	\begin{subfigure}[b]{0.48\textwidth} 
		\centering
		\includegraphics[width=1\textwidth]{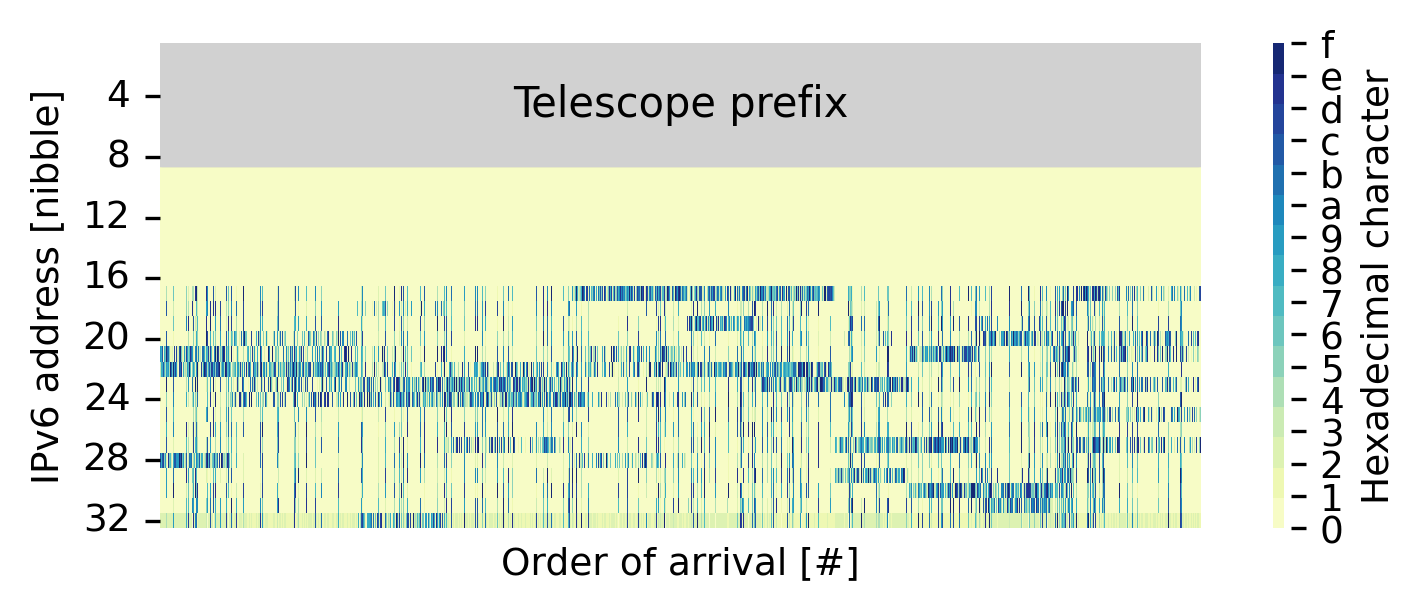}
		\caption{Structured}
		\label{fig:behavior:tencent}
	\end{subfigure}\hfill
	\begin{subfigure}[b]{0.48\textwidth} 
		\centering
		\includegraphics[width=1\textwidth]{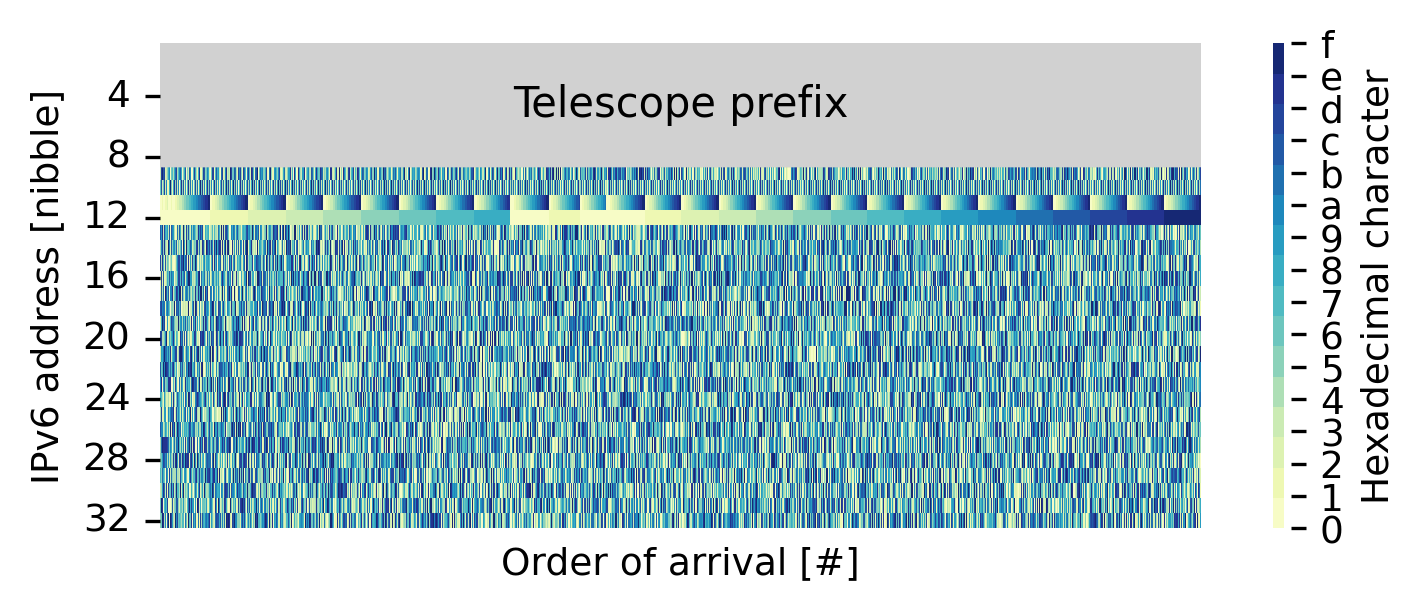}
		\caption{Random}
		\label{fig:behavior:ponynet}
	\end{subfigure}\hfill
	\caption{Structured and randomized target address generation observed in two scan sessions. Sequentially targeted destination addresses are represented in hex using a different color for each digit value.}
	\label{fig:behavior:hex_plot}
\end{figure}

\begin{figure}
	\begin{minipage}[t]{0.48\linewidth}
		\centering
		\footnotesize
		\vfill
		\includegraphics[width=1\textwidth]{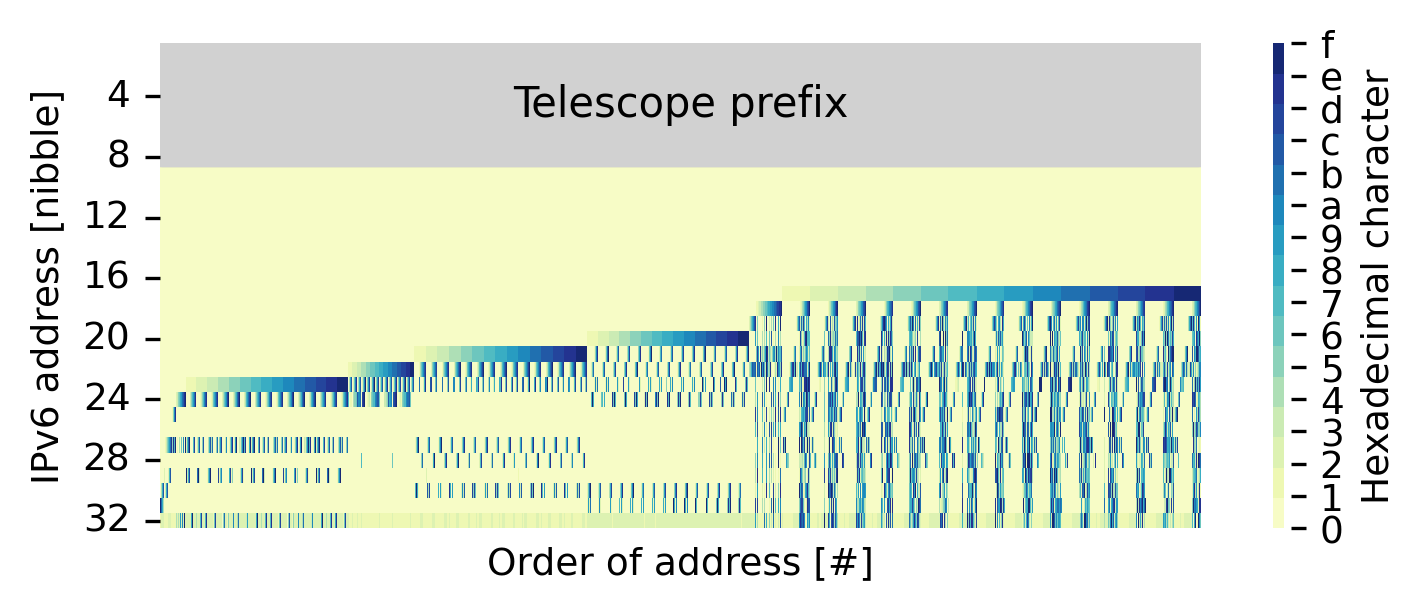}
		\caption{Addresses from Fig.~\ref{fig:behavior:tencent} numerically sorted.}
		\label{fig:behavior:sorted}
	\end{minipage}\hfill
	\begin{minipage}[t]{0.48\linewidth}
		\centering
		\footnotesize
		\vfill
		\vspace{5pt} 
		\includegraphics[width=1\textwidth]{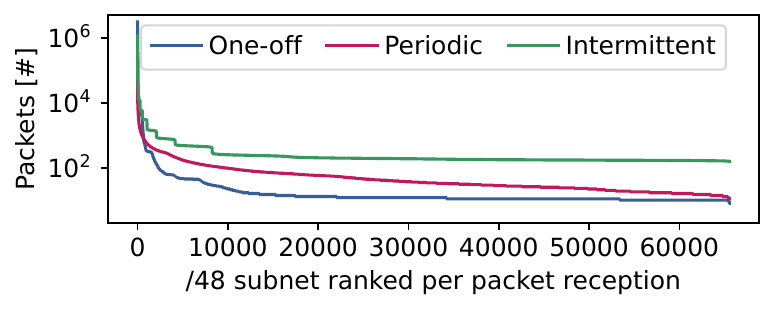}
	\caption{Packets per scanner type  across /48~subnets.}
	\label{fig:behavior:subent-coverage-overview}
	\end{minipage}
\end{figure}

\paragraphNoDot{Address selection of scanners}
 classifies into the categories \one structured, \two random, and \three unknown according to our taxonomy (\cf \autoref{sec:taxonomy}).
\autoref{fig:behavior:hex_plot} illustrates samples of a structured and a
 random selection  session.
For each scan session, we show all targeted addresses in hexadecimal representation  (y-axis) from low nibble (top) to high nibble (bottom) and rank them by time of arrival. Our telescope prefix is concealed in gray.
\autoref{fig:behavior:tencent} illustrates a scan session by AS132203 with 151k packets. %
While  most nibbles across the figure are zero, periods with frequently changing characters are visible as stripes.
By sorting the addresses lexicographically,  \autoref{fig:behavior:sorted} visualizes a clear structure of the traversal through the network.
Although blocks of zeros become more pronounced, iterative traversal is now visible in stripes of shifting color (from \texttt{0} to \texttt{f}).
After roughly two thirds of the packets the pattern shifts and shows a tree like structure from traversing into the `leaves' of some subnets.
\autoref{fig:behavior:ponynet} depicts a scan session by AS53667 with 113k packets. %
Nibbles 11 and 12 show a structured iteration through our subnets, but there is no pattern or structure apparent in the remaining segments.
This observation suggests a random generation of the last 80 bits of the target addresses. \autoref{appendix:nistresults} presents a randomness assessment of the IID and subnet parts, using the test suite from the National Institute of Standards and Technology (NIST)~\cite{brsns-stsrp-10}.

The prevalent probing strategy within scan sessions is for structured addresses (see \autoref{fig:behavior:taxonomy_t1}). Still, many sessions randomly traverse the address space, especially those from periodic scanners. While random probing can be useful to detect aliased prefixes, it is commonly used in topology measurements to detect routers~\cite{b-yirhs-16,bdpr-ibsai-18,yc-eirid-24,rb-dtnp-20,llzdl-finpd-21,hmu-drwie-24} as it is likely to hit unassigned addresses or networks, which triggers ICMP error messages from on-path routers and reveals network topology.


\paragraph{Temporal behavior of scanners}
Our classification of temporal behavior as summarized in Table~\ref{tbl:behavior:source_classification} shows that the vast majority of scanners appear only once (69.7\%), which may relate to our limited observation period between the biweekly announcements. Intermittent and periodic scanners remain with a similar share around 15\%.
Scanners of periodic behavior (14.8\%) comprise the majority of sessions (72.8\%) due to short periods.
One-off sessions contain fewer packets and are rather short, \ie always less than 4 hours (top 10 $\geq$ 1 hour each). These scanners may act very coarse-grained or use a distributed scan infrastructure, which requires little work from individual sources. The latter could show up as comparatively many associated scanning source addresses with fewer packets each. In our dataset, we detect such behavior from RIPE Atlas probes and Alpha Strike Labs.
In contrast, scan sessions from periodic and intermittent scanners tend to be much longer. The longest session from a periodic scanner spans over 311 hours (top 15 $\geq$ 78 hours each); for intermittent, the longest session spans 316 hours (top 15 $\geq$ 6 hours each).

Beyond session schedules, the coverage of the /48 subnets in our /32 telescope prefix is of interest. We observe two kinds of behavior, \one probing a wide range of /48 subnets with only a few packets each, or \two focusing on a few subnets and probing these in more depth. \autoref{fig:behavior:subent-coverage-overview} ranks all subnets from highest to lowest packet reception.  Intermittent scanners probe the majority of subnets more evenly, while one-off scanners focus on a few selected subnets. Periodic scanners probe a wider range of subnets with more packets but tend to visit subnets selectively. Periodic and intermittent scanners focus on the \texttt{0000} subnet while one-off scanners probe the \texttt{e000} subnet the most.

\paragraph{Cross-category behavior}
All three temporal behaviors show specific patterns w.r.t. the target network and address selection.
One-off sessions are highly biased towards single-prefix scans (95\%) and often follow a structured target selection approach.
In contrast, sessions of intermittent scanners predominantly show inconsistent behavior (50\%) followed by single-prefix scanning (32\%).
While we also observe most periodic sessions as inconsistent (54\%), 39\% scan independent of network-size. Note that in terms of individual scan sources we observe inconsistent behavior for only 13 intermittent (0.7\%), and 51 periodic scanners (2.9\%).  
The results of address selection, temporal behavior, and network target selection provide different perspectives on the observed scanning behavior. Our experiment clearly reveals a trend towards structured, BGP-aware scanning.

\subsection{Scan Tools and Scanning Sources}
\label{sub:scan-tools}
We analyze all scan sources and packet payloads (if present) for patterns that allow for a better understanding of the scanners that visit our telescope, including the tools they use.

\setlength{\tabcolsep}{3.00pt}
\begin{wraptable}{r}{0.5\textwidth}
	\footnotesize
	\centering
      \caption{Overview of identified scan tools. RIPE Atlas probes account for 55\% of all sources we observe at T1.}
	\label{tbl:scantools:overview} 
	\vspace{-10pt}
		\begin{tabularx}{0.5\columnwidth}{Xrrrr}
			\toprule
			& \multicolumn{2}{c}{Scanners} & \multicolumn{2}{c}{Sessions} \\ 
			\cmidrule(lr){2-3}
			\cmidrule(lr){4-5}
			Scan Tool & \makecell[c]{[\#]} & \makecell[c]{[\%]} & \makecell[c]{[\#]} & \makecell[c]{[\%]}  
			\\
			\midrule
      RIPEAtlasProbe~\cite{ripeatlasprobe}& 6483 & 54.82 & \num{11859} & 12.87 \\
      Yarrp6~\cite{yarrp} & 22 & 0.19 & 562 & 0.61 \\
      Traceroute~\cite{traceroute} & 19 & 0.16 & 163 & 0.18 \\
      Htrace6~\cite{htrace6} & 9 & 0.08 & 16 & 0.02 \\
      6Seeks~\cite{6seeks} & 5 & 0.04 & 17 & 0.02 \\
      6Scan~\cite{6scan} & 3 & 0.03 & 17 & 0.02 \\
      CAIDA Ark~\cite{caidaark} & 2 & 0.02 & 2019 & 2.19 \\
			\bottomrule
		\end{tabularx}
\end{wraptable}
\paragraph{Public tools used by scan sources}
17M (40\%) of 43M captured packets contain a payload. These packets are sent by 11,001 (93\%) scan sources and cover $\approx 70K$ (76\%) of all sessions. We analyze the  payloads for fingerprints and map them to public tools.
 We were able to identify eight public tools (see \autoref{tbl:scantools:overview}), mostly traceroute-type implementations for IPv6 topology and periphery measurements. Interestingly, we observe scanners using the \textit{Htrace6} tool in December 2023,
 even though the code was first published in late January 2024. %
The traceroute-type tool \textit{Yarrp6} is used by 22 distinct scan sources, all appear periodically.
It is the only open source tool which we observe regularly over the complete observation period.
A large fraction of all scan sources~(55\%) are RIPE Atlas probes, almost exclusively identified as one-off scanners. They always target the \texttt{::1} addresses in each prefix.

\paragraph{Context of scan sources}
While the majority of scan sources (96\%) originate from hosting or ISP networks (see \autoref{tbl:scanorigin:overview}), most of them can be associated with a research context -- 97\% of the sources in ISP networks and 22\% in the hosting networks belong to RIPE Atlas probes.  58\% of  sources in the hosting category belong to a single company (Alpha Strike Labs) that commercializes research scanning. 
Heavy hitters contribute a significant portion of all packets, but only account for a negligible fraction of sessions (\cf \S\ref{sec:traffic}). \autoref{tbl:scanorigin:overview} shows that three out of four heavy hitters are located in hosting networks, without research context, thus indicating malicious activity.

Our findings emphasize that research scanners probe the IPv6 address space more often and regularly, making them a good reference point for telescope setup as they predominantly act BGP-aware. Furthermore, we find 18 scan sources to live-monitor BGP announcements, since we reliably observe them within 30 minutes after a new BGP announcement. 

\begin{wraptable}{r}{0.5\textwidth}
	\footnotesize
		\centering
		\vspace{-8pt}
    \caption{Network types of scan sources. Most sessions (76\%) and sources (96\%) originate from hosting and ISP networks. Three out of four heavy hitters (Hit.) are located in hosting networks.}
  \label{tbl:scanorigin:overview}
  \vspace{-10pt}
		\begin{tabular}{lrrrrrr}
			\toprule
      & \multicolumn{2}{c}{Scanners} & \multicolumn{2}{c}{Sessions} & \multicolumn{2}{c}{Packets}\\ 
			\cmidrule(lr){2-3}
			\cmidrule(lr){4-5}
			\cmidrule(lr){6-7}
      Network & \makecell[c]{[\#]} & \makecell[c]{[\%]} & \makecell[c]{[\#]} & \makecell[c]{[\%]} &  \makecell[c]{[\#]} & \makecell[c]{[\%]}
	  \\
			\midrule
Hosting & \num{6624} & 56.01 & \num{23682} & 25.70 & \num{28371475} & 65.06 \\
\multicolumn{1}{r}{w/o Hit.}& \num{6621} & 55.99 & \num{23674} & 25.69 & \num{4496454} & 10.31 \\
ISP & \num{4681} & 39.58 & \num{46864} & 50.85 & \num{1478591} & 3.39 \\
Education & \num{245} & 2.07 & \num{17634} & 19.14 & \num{13629270} & 31.25 \\
\multicolumn{1}{r}{w/o Hit.} & \num{244} & 2.06 & \num{17627} & 19.13 & \num{4375030} & 10.03 \\
Business & \num{194} & 1.64 & \num{2259} & 2.45 & \num{71689} & 0.16 \\
Government & \num{6} & 0.05 & \num{7} & 0.01 & \num{96} & 0.00 \\
Unknown & \num{76} & 0.64 & \num{1707} & 1.85 & \num{58527} & 0.13 \\
\bottomrule
		\end{tabular}
\end{wraptable}
\paragraph{Source overlap across telescopes}
Only one /128 scan source, originating from a hosting network, probed all four telescopes during the initial 12 weeks, at the end of September and early October.
These probes matched the signature of ``Yarrp6''.
The same scan source returned to T2 in November, but with a different signature.
Either the same entity ran a new scan campaign or a different entity incidentally used the same IP address.

\begin{figure}
	\centering
	\begin{subfigure}{0.48\textwidth}
		\centering
		\includegraphics[width=1\textwidth]{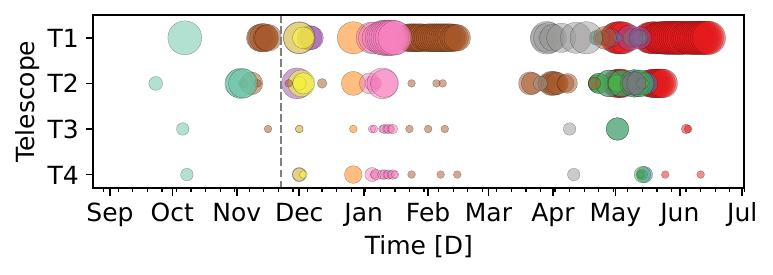}
	  \caption{Daily activity from scan sources across all four  telescopes. Bubble sizes signify packet counts and colors represent scan sources (same color means same scan source).}
		\label{subfig:overlap:packets}
	\end{subfigure}
	\hfill
	\begin{subfigure}{0.48\textwidth}
		\centering
		\includegraphics[width=1\textwidth]{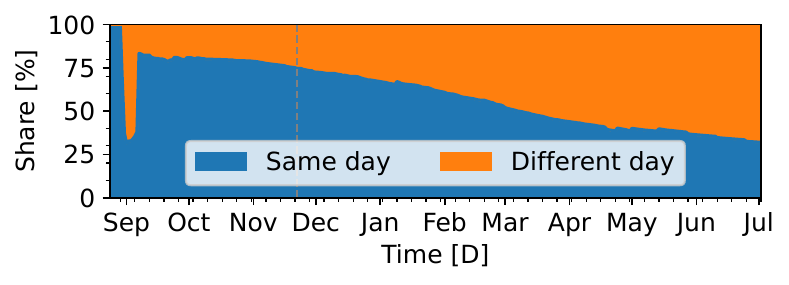}
		\caption{Cumulative share of sources observed in T1 and T2 on the same or different days.\\\\}%
		\label{subfig:overlap:difference}
	\end{subfigure}
	\vspace{-4pt}
	\caption{Analysis of overlapping sources. The dashed line marks the start of our active experiment.}
	\vspace{-16pt}
\end{figure}

Nine other /128 scan sources were observed at every telescope during the entire measurement period. \autoref{subfig:overlap:packets} visualizes the daily activities, marking each scan source with a distinct color.
Larger markers signify more packets on a given day.
For each source, T1 and T2 together received about 98\% of the packets, although in most cases one of them was focused ($\ge$90\%).
While most scan campaigns were localized in time, some scan sources showed up repeatedly across a subset of the telescopes.
As an example, in mid-November one source was observed in T1, T2, and T3, but not in T4.
Overall, six of these ten~scan sources belong to four different hosters.
The remaining four scan sources are located in three research networks.
Two of them were both active in May, largely overlapping in time, and were likely part of the same scan campaign.

Limiting the analysis to our separately announced prefixes (T1 and T2) reveals an overlap of 2,169 addresses in total or 598 addresses during the initial period.
\autoref{subfig:overlap:difference} shows the cumulative share of scan sources observed on the same day in blue and on different days in orange.
After few addresses were observed on the first days, the dip in September followed by a sudden rise was caused by 254 new sources scanning on the same day.
Roughly 75\% of scan sources were observed in both telescopes on the same day during the initial period.
Attracting scanners by our active experiment  only in T1 leads to a drifting apart between the telescopes as the share declines to approximately 30\% over the course of our active experiment.

\paragraph{Correlation with the TUM hitlist}
While our prefixes appear on the TUM hitlist within a few days after the announcement, there is no noticeable impact on the number of sessions or traffic, implying that the prefixes on the TUM hitlist are rarely used by BGP-reactive scanners.

\paragraph{Key observations} Announcing more specific prefixes in BGP attracts significantly more traffic (+286\%) and scanners (+275\%), which in contrast to stable prefixes prefer one-time visits (69.7\%) and single-prefix scans. A large portion of scanners probes prefixes independent of its size and applies a structured target selection. The majority of scanners can be tied to its origin or a public tool.

\section{Discussion, Conclusion, Outlook}
\label{sec:conclusion}
\label{sec:discussion}

In this paper, we explored the behavior of IPv6 scanners from the perspective of four network telescopes with contrasting properties.
Our telescopes were passive, traceable, or (re-)active.
To specifically study scanners that are triggered by BGP signals, we introduced and utilized a controlled, active measurement method that relies on IPv6~prefixes of varying sizes announced proactively~in~BGP.

\paragraphNoDot{What should an IPv6 telescope operator consider?} Our findings include the following practical implications for operating network telescopes. \one Visibility of a network largely increases if its prefix is individually announced in BGP instead of being a subnet that is only part of a covering prefix; \two the size of an IPv6 prefix is of lower relevance for a network telescope than the number of individually announced prefixes; \three different attractors, \eg DNS versus BGP, draw different kinds of scanners; \four active network services draw scanners to neighboring address space; \five structured target addresses are preferred by many scanners.

\paragraphNoDot{Are  observations in telescopes unbiased?}
No. Scanners seem to contact telescopes following external triggers from the network or the application side, which in turn means that triggers attract only those scanners that react to them. We are aware of the different specific biases related to our network telescopes and emphasize our converse finding: We measure the effects of network triggers and show, how and which scanners react to them, \ie we quantify the biasing factors.

\paragraphNoDot{Are IPv6 telescopes suitable to monitor DDoS?}
No. Telescopes commonly monitor DDoS by capturing the backscatter from randomly spoofed attack traffic. It is very unlikely to capture packets with randomly selected IPv6 destination addresses in a telescope. Researchers and security experts will need to identify new ways to assess DDoS---without relying on IPv6 background radiation.

\paragraphNoDot{What next?} This work gives rise to the following research directions.
\one Future measurements and analyses shall quantify the effect of further triggers that attract traffic to IPv6 network telescopes.
\two With various, compatible trigger measurements at hand, a correlation analysis should enable a more realistic, quantitative assessment of the biases inherited from IPv6 for network telescopes.  
\three New  methods for Internet observatories  are needed to capture IPv6 background radiation.

\begin{acks}
This work was partly supported by the \grantsponsor{BMFTR}{Federal Ministry of Research, Technology and Space (BMFTR)}{https://www.bmbf.de/} within the projects IPv6Explorer (\grantnum{BMFTR}{16KIS1815}) and AI.Auto-Immune (\grantnum{BMFTR}{16KIS2332K} and \grantnum{BMFTR}{16KIS2333}).
\end{acks}

\label{lastbodypage}

\bibliographystyle{ACM-Reference-Format}
\bibliography{own,internet,security,rfcs,big-data,theory}

\appendix

\section{Ethics}

This work does not raise any ethical issues.

\section{Artifacts}
All artifacts of this paper are publicly available. These include \one all raw measurement data that we captured during our 11 month observation period; \two all data derived from postprocessing \ie data that substantiate our arguments and serve as input for our figures; \three post-processing scripts. All artifacts and details on how to use them are archived here: \url{https://doi.org/10.5281/zenodo.16419096}.

\section{Testing for randomness: NIST Test Suite}
\label{appendix:nistresults}
The NIST Test Suite comprises 15 different randomness tests.
We exclude tests that either require over 1,000 bits of randomness or additional information. %
This leaves us with four tests for our analysis.

\paragraphNoDot{Frequency (monobit)} tests the balance between ones and zeros in a sequence to determine uniform  randomness.
Failure in this initial test make other tests likely to fail as well.

\paragraphNoDot{Runs}
examines the number of uninterrupted sequences of identical bits to evaluate the randomness of the oscillation between ones and zeros.

\paragraphNoDot{Discrete Fourier transform (spectral) (FFT)}
analyzes peak heights in the Discrete Fourier Transform of a sequence to identify (non-random) periodic features.

\paragraphNoDot{Cumulative sums (cusum)}
evaluates deviation of the sum of numbers from the expected values of a random sequence.
It can be applied forward (cusum0) or backward (cusum1).

\paragraph{NIST test input}
We selected sessions with at least 100 packets from our dataset.
This filters out all but 2219 sessions, roughly 2.4\%, which include 94\% of all packets.
Since we observed scanners that use different approaches for different parts of the address, we test address sections separately: the first 32 bits after our fixed /32 prefix (subnet) and the last 64 bits (interface identifier (IID)).

\begin{figure}%
	\centering
	\includegraphics[width=0.8\textwidth]{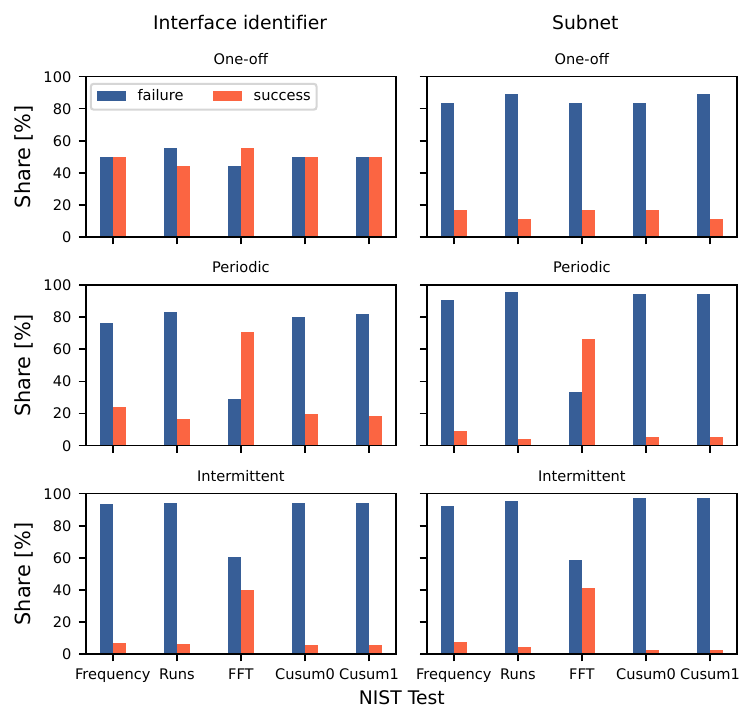}
	\caption{Results of the NIST tests for T1 sessions with $\ge$ 100 packets. Scanners tend to iterate IIDs more randomly (\ie test success) compared to subnets.}
	\label{fig:behavior:nist}
\end{figure}

\paragraph{NIST test results}
\autoref{fig:behavior:nist} shows the share of scan sessions in our data set that succeed and fail in the selected NIST tests, separated by IID (left) and subnet (right).
Results are further categorized by the temporal behavior of the scanner.
Success signifies the selection to be random.
For subnets, NIST test mostly fail (right column in \autoref{fig:behavior:nist}).
In contrast, IID selections pass more frequently the NIST test than selected subnets.
Combining both observations suggest that scanners favor a structured approach to select subnets but are more likely to choose random addresses inside the prefixes.
These results bolster our observations in~\autoref{fig:behavior:hex_plot}.

Non-random selection of IIDs occurs most often in \textit{periodic} and \textit{intermittent} scanners.
Their repeated scanning cycle might influence their selection as they want to get a comprehensive view of the address space instead leaving it to chance.
Interestingly, \textit{one-off} scanners are more likely to randomly select IIDs than the other categories.
However, no category shows scans with predominantly random selection.

We do not know how TGAs influence the selection of addresses and how they influence the outcome of this analysis.
This will be part of our future work.
These results support our assumption.
Random scanning cannot be detected by categorizing targeted addresses individually but requires context from the scan session.

\end{document}